\documentclass[useAMS,usenatbib]{mn2e}
\usepackage{aas_macros}
\usepackage{amssymb}
\usepackage{setspace}
\usepackage[pdftex]{graphicx}
\usepackage{comment}
\usepackage{url}

\newcommand {\apgt} {\ {\raise-.5ex\hbox{$\buildrel>\over\sim$}}\ }
\def\eg{{\it e.g.\ }}
\def\ie{{\it i.e.\ }}

\def\bellcite{B02\ }
\def\bellcitedot{B02.\ }
\def\bellcitecom{B02,\ }

\title[Effective viscosity from cloud-cloud collisions in three-dimensional global SPH simulations]{Effective viscosity from cloud-cloud collisions in three-dimensional global SPH simulations}
\author[D. J. Williamson and R. J. Thacker]{D. J. Williamson$^{1}$\thanks{E-mail:
williams@ap.smu.ca} and R. J. Thacker$^{1}$\\
$^{1}$Department of Astronomy and Physics, St Mary's University, Halifax B3H 3C3, Canada\\}
\begin{document}

\date{This pre-print submitted Jan 2012 (Accepted to MNRAS, Dec 2011) }

\pagerange{\pageref{firstpage}--\pageref{lastpage}} \pubyear{2011}

\maketitle

\label{firstpage}

\begin{abstract}

Analytic estimates of the viscous time-scale due to cloud-cloud collisions have been as high as thousands of Gyr. Consequently, cloud collisions are widely ignored as a source of viscosity in galactic disks. However, capturing the hydrodynamics of discs in simple analytic models is a challenge, both because of the wide dynamic range and importance of 2D and 3D effects. To test the validity of analytic models we present estimates for the viscous time-scale that are measured from three dimensional SPH simulations of disc formation and 
evolution. We have deliberately removed uncertainties associated with 
star-formation and feedback thereby enabling us to place lower bounds on 
the time-scale for this process. We also contrast collapse simulations 
with results from simulations of initially stable discs and examine 
the impact of numerical parameters and assumptions on our work, to 
constrain possible systematics in our estimates. We find that 
cloud-collision viscous time-scales are in the range of $0.6$--$16$ Gyr, considerably shorter than 
previously estimated. This large discrepency can 
be understood in terms of how the efficiency of collisions is included 
in the analytical estimates. We find that the viscous time-scale only depends weakly on the number of clouds formed, and so while the viscous time-scale will increase with increasing resolution, this effect is too weak to alter our conclusions.

\end{abstract}


\section{Introduction}

Arguably, the most successful model for the formation 
of disc galaxies is the $\Lambda$CDM model, in which galaxies are formed 
from the dissipational collapse of baryonic gas within a dark matter 
halo \citep{1978MNRAS.183..341W,1984Natur.311..517B,1985ApJ...292..371D,1991ApJ...379...52W,1993MNRAS.264..201K,1994MNRAS.271..781C,2006RPPh...69.3101B,2010PhR...495...33B}. While the 
physical viscosity of the baryonic gas is not anticipated to have 
a strong influence on gas evolution except in magnetized or hot 
environments such as a galaxy cluster 
\citep{2006MNRAS.371.1025S}, 
effective kinematic viscosities could in principle impact disc 
evolution. Simulations by \citet{1987ApJ...320L..87L} with a viscous 
time-scale close to the star-formation time-scale showed that viscous 
evolution with infall can reproduce the ubiquitous exponential 
density profile from a range of initial conditions. In this work the 
viscosity was assumed to be caused by large-scale turbulent motions 
dissipating kinetic energy and transporting angular momentum.

Feedback from supernovae can be a source of viscosity by feeding this 
turbulence \citep{2003A&A...404...21V}. Additionally, the self-gravity of the gaseous 
disc can 
provide an effective viscosity \citep{2002A&A...382..872V}. This can take the form of  
large-scale instabilities \citep{2009ApJ...704..281R,2001ApJ...553..174G}, 
or of interactions 
between giant molecular clouds \citep{2002A&A...382..872V}. These cloud interactions potentially 
generate 
viscosity through two different mechanisms. Firstly, gravitational 
scattering can increase the velocity dispersion of the cloud population, 
converting orbital energy into large-scale turbulence \citep{1991ApJ...378..565G,1989PASJ...41..241F,2009MNRAS.392..294A}. 
Secondly, during inelastic 
collisions between clouds, shocks convert orbital energy into turbulence 
and heat within the colliding clouds \citep[\eg][]{2003MNRAS.340..841G,2007MNRAS.378..507K,2009A&A...504..437A}. Radiative processes contribute to 
the dissipation of kinetic energy during these collisions, and are also 
important for dissipating turbulent energy that has cascaded into 
thermal energy. These processes are significant even in the absence of 
star-formation: the observations compiled by \citet{2006ApJ...638..797D} 
show that the velocity dispersion of H{\sc I} gas does not strongly 
depend on the star-formation rate below a certain threshold, and the 
AMR simulations of \citet{2009MNRAS.392..294A} suggest that a `baseline' 
turbulence is caused by interactions between clouds, and that this is 
only supplemented by supernova feedback at high star-formation rates.

It has been argued \citep[][hereafter B02]{2002A&A...382..872V,2002ApJ...581.1013B} that 
cloud-collisions are not an efficient source of viscosity. In 
particular, in \bellcite the time-scales 
for viscosity due to cloud collisions are estimated to be on the order of $t_\nu\sim 
1000$ Gyr in most local spiral galaxies, although the time-scales might 
be considerably lower in earlier gas-rich galaxies or in galaxies where 
the velocity distribution of GMCs has been stirred up by some mechanism 
\citep[such as galaxy interaction \eg][]{1995ApJ...448...41H}. \citet{2002A&A...382..872V} 
argues that 
because molecular clouds evaporate at an age of $\sim10^7$ yr, and this 
is less than the time between collisions ($\sim 10^8$ yr), cloud 
collisions are very rare. However, cloud formation times, assuming that 
collapse and 
formation of $H_2$ are the dominant factors in forming a cloud, appear 
to 
equally short \citep{2007ApJS..169..239G}. This 
leads to a scenario in which the number density of clouds is roughly 
constant, although the short
life-time may affect the velocity dispersions of molecular clouds as
they have less time to build up a large deviation from circular velocity
through scattering events with other clouds. In this steady state, the effective collision 
time-scale should remain similar.

It has also been argued 
that physical collisions between clouds have a smaller effect than 
gravitational scattering \citep{1988ApJ...328..404J}, particularly as if 
magnetic fields are taken into account, cloud interaction cross-sections 
may be underestimated \citep{1998A&A...337..105O}. On the other hand, 
\citet{1996ApJ...462..309D} modelled a system of cloud particles, 
finding that cloud collisions rather than local gravitational 
interactions (scattering events) dominate the mass distribution and 
velocity dispersion of molecular clouds, suggesting that cloud 
collisions may indeed be important. However, as far as we are aware, the 
effective viscosity of direct 
cloud-cloud collisions has not yet been examined in global three 
dimensional numerical hydrodynamic models.

Most simulations of cloud formation and the associated disc dynamics 
have been performed in two dimensions and/or on a small scale using 
shearing-box studies \citep[\eg][]{2007ApJ...660.1232K}. However, 
increased computing power and the availability of locally adaptive 
algorithms have recently enabled galaxy-scale simulations with sufficiently high 
resolution to resolve cloud-formation in discs. Numerical experiments have been performed using 
both AMR \citep{taskertan,2011ApJ...730...11T,2009MNRAS.392..294A} and SPH 
\citep{robertsonkravstov} with resolutions as fine as $6$ pc. The 
non-trivial cooling processes and chemistry make these simulations a 
significant technical challenge. \citet{2009MNRAS.392..294A} and 
\citet{taskertan} ran suites of high resolution AMR simulations of 
Milky-Way and M33-like disc galaxies, and reported on the properties of 
the clouds generated by their models, including cloud-cloud velocity 
dispersion. However, neither study has provided an estimate of the 
viscous time-scale due to cloud-cloud collisions. Furthermore, the discs 
of \citet{taskertan} are much more stable than the Milky Way, with a 
density distribution chosen to give a constant value of the Toomre Q 
parameter \citep{1964ApJ...139.1217T}, 
and a static dark matter and stellar component, which may 
inhibit some of the instabilities important to cloud formation.

In this paper we revisit the calculations of \bellcite 
with full three dimensional SPH models. This is not entirely trivial 
since there is no universally agreed upon cloud finding process. 
However, the use of a particle method enables the Friends-of-Friends \citep{1985ApJ...292..371D}
group finding methodology and we adapt that to our simulations. Hence 
given our cloud population our primary goal is to see whether the 
analytical calculations are supported, and if not what the implications 
are. It is important to 
note that the results of such simulations could 
highlight non-physical evolution in numerical schemes with artificial 
viscosities, of which SPH is a notable example \citep{2011A&A...526A.158V}. We also investigate the 
issue of numerical artefacts in our calculated results. This is a key 
issue since structure formed within simulations starting from smooth 
initial conditions is inevitably the result of amplification of noise in 
the initial conditions.

While a full calculation in the cosmological context \citep[\eg][]{1992ApJ...391..502K,2001ApJ...555L..17T,2007MNRAS.374.1479G,2010MNRAS.408..812S,2009MNRAS.396..696S,2008ApJ...689..678B}
is beyond the scope of this paper, primarily due to resolution 
limitations, we instead 
consider two classes of isolated models. We examine an equilibrium 
system with similar parameters to the Milky Way consisting of a gas disc, a stellar disc and bulge, and a dark 
matter halo. Here the gas disc 
is stabilized by the other components which dominate the system's mass. 
We also consider the dissipational collapse problem that has been used 
extensively elsewhere \citep[\eg][]{1976ApJ...204..649G,1984ApJ...286..403C,1991ApJ...377..365K,2004MNRAS.349...52B,2006MNRAS.370.1612K}. By 
contrast with the 
Milky Way model, this collapse produces a very unstable disc, and so we 
investigate both high-stability gas-poor systems and 
low-stability gas-rich systems. These models include hydrodynamics, 
gravitational interactions, and cooling with a dynamic temperature 
floor. By removing the numerous unknowns associated with star-formation 
and feedback \citep[as discussed in numerous 
places \eg][]{2000ApJ...545..728T,2009ApJ...695..292C,2010ApJ...717..121C} we hope to 
isolate the impact 
of cloud-cloud interactions and place lower bounds on the viscous 
time-scale.

The layout of the paper is as follows: in section \ref{sim} we outline 
the details of our simulation code. We also discuss the initial 
conditions, our cloud identifying approach and also the underlying 
theory of the effective viscosity. Results are presented in section 
\ref{results} followed by a brief conclusion.

\section{Simulation}\label{sim}

\subsection{Simulation Code}

We model the dark matter, stars and gas using a specially adapted version of 
the OpenMP n-body AP$^3$M \citep{ap3m} SPH \citep{1992ARA&A..30..543M} 
code HYDRA \citep{phydra}. Our modifications are:

\begin{figure}
\begin{center}
\includegraphics[width=1.\columnwidth]{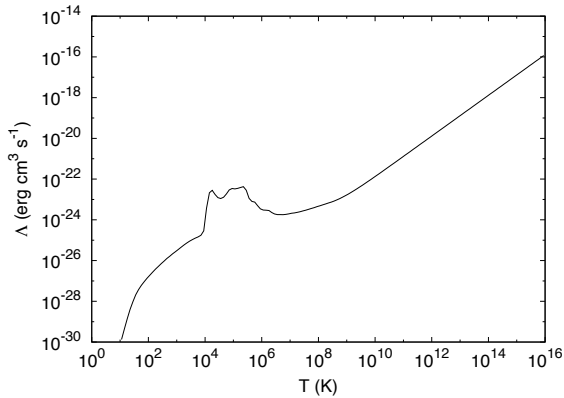} 
\end{center}
\caption{\label{lambdaplot} 
The cooling curve used in our models. Values below $10^4$ K are from \citet{WadaNorman}, while those above $10^4$ K are from \citet{1993ApJS...88..253S}.}
\end{figure}

\begin{enumerate}
\item
The cooling curve has been extended to down to $10$ K using the cooling function 
($\Lambda$) of \citet{WadaNorman}, although we set our fiducial 
temperature floor to 300K to make our results more comparable with 
\citet{taskertan}, except in cases where we investigate the effect of a 
lower floor. The earlier cooling curve of \citet{1993ApJS...88..253S} is retained for $T>10^4$ K. The combined cooling curve is plotted in Fig.~\ref{lambdaplot}.

\item
We implemented a dynamic temperature floor based on \citet{robertsonkravstov}, described in section \ref{tfloor}.

\item
The parallelisation algorithm has been altered so that
during the particle-particle gravity and SPH calculation regions 
containing a large number of particles are split over all processors, instead of each 
processor receiving a single region. This greatly improves load-balance 
in simulations containing many dense clumps of particles.

\end{enumerate}

\subsubsection{Dynamic Temperature Floor}\label{tfloor}
We use a method similar to \citet{robertsonkravstov} to ensure that the 
Jeans mass is resolved in our simulations. This is to satisfy the 
\citet{1997ApJ...489L.179T} criterion and avoid artificial fragmentation 
- crucial in simulations of cloud formation. The method is in the form 
of a dynamic pressure floor. The Jeans mass\citep{1902RSPTA.199....1J}  is defined 
as
\begin{equation}
m_{\rm Jeans}=\frac{\pi^{5/2}c_s^3}{6G^{3/2}\rho^{1/2}}.
\end{equation} 

\citet{1997MNRAS.288.1060B} noted each 
particle should satisfy $2N_{neigh}m_{gas}<m_{\rm Jeans}$ (where 
$N_{neigh}$ is the number of SPH neighbours for the particle and 
$m_{gas}$ is the gas particle mass) to avoid artifical fragmentation. 
Defining the local ratio of the Jeans mass to the SPH 
kernel mass as $h_{\rm Jeans}$, then this requirement is equivalent to the 
statement that
\begin{equation}
h_{\rm Jeans}=\frac{\pi^{5/2}c_s^3}{6G^{3/2}N_{neigh}m_{gas}\rho^{1/2}}<N_{\rm Jeans}
\end{equation} with $N_{\rm Jeans}$ being the required factor by which the Jeans mass must be resolved, which in the Bate and Burkert case is set to $2$. In an ideal gas 
$c_s\propto\sqrt{u}$, so we can fulfill this criterion by applying
\begin{equation}
u\rightarrow u\times\left(\frac{N_{\rm Jeans}}{h_{\rm Jeans}}\right)^{2/3}
\end{equation} whenever $h_{\rm Jeans}<N_{\rm Jeans}$.

We found spurious (sub-resolution) string-like structures forming within clouds for low values of $N_{\rm Jeans}$, and found that $N_{\rm Jeans}=50$ removed these structures and resulted in a more homogeneous interior for clouds.

\subsection{Initial Conditions}
\subsubsection{Milky Way Model}\label{mwics}

We produce our Milky Way model using the GALACTICS package \citep{1995MNRAS.277.1341K, 2005ApJ...631..838W, 2008ApJ...679.1239W} with the parameters in Table 2 of \citet{2008ApJ...679.1239W}. Through an iterative process, this package produces an equilibrium system consisting of an exponential stellar disc, a stellar bulge and a dark matter halo. The disc is exponential radially, follows sech$^2$ vertically, and has a radial dispersion profile of
\begin{equation}\sigma^2_R(R)=\sigma^2_{R0}\exp(-R/R_\sigma),\end{equation} 
where we set $R_\sigma=R_d$ for simplicity. We generate the gas disc by copying 
the disc star particle positions and flipping the coordinates across the $x=y$ plane to prevent particles having coincident positions. Bulge particles are not copied. The masses of the gas and star particles are scaled to give the appropriate mass ratio. The gas disc is given a dispersionless velocity profile output by GalactICs and is initially isothermal at $10^{4}$K. The disc scale length is $2.81$ kpc, truncated at $30$ kpc with a truncation scale-length of $0.1$ kpc. The scale height is initially $0.36$ kpc, and the total disc mass is $5.2\times10^{10}M_\odot$.
The halo density profile is
\begin{equation}\label{eqhalodensity}
\rho =\frac{2^{2-\gamma}\sigma_h^2}{4\pi a_h^2}\frac{ 1}{(r/a_h)^\gamma (1+r/a_h)^{3-\gamma}} \frac{1}{2} \mathrm{erfc} \left( \frac{r-r_{h}}{\sqrt{2}\delta r_h} \right),
\end{equation} where $a_h$ is the halo scale parameter, $r_{h}$ is the cutoff radius, $\delta r_h $ is the scale length for the truncation, $\gamma$ is the `cuspiness' parameter (equal to unity for an NFW profile), and $\sigma_h$ is a velocity parameter that sets the mass of the halo. 

\begin{table*}
\begin{tabular}{ c | c | c | c | c | c | c |}
\hline\hline
Name & $a_h$ (kpc) & $r_h$ (kpc)  & $\delta r_h$ (kpc) & $\gamma$ & $\sigma_h$ (km s$^{-1}$) & $M_{halo}$ \\
\hline
Collapse haloes & $25.75$ & $300$ & $50$ & $1.0$ & $351$ & $1.1\times10^{12}$\\
MW haloes & $13.6$ & $275$ & $25$ & $0.81$ & $330$ & $7.3\times10^{11}$\\
\hline
\end{tabular}
\caption{\label{haloprams}Halo parameters. As in Eq.~\ref{eqhalodensity}, $a_h$ is the halo scale parameter, $r_h$ is the truncation radius, $\delta r_h$ is the scale length for this truncation, $\gamma$ is the cuspiness parameter, and $\sigma_h$ is a velocity parameter that sets the halo mass, $M_{halo}$.} 
\end{table*}

Halo parameters are given in Table~\ref{haloprams}. We also ran a test with a static analytic halo potential, to explore if the discretization of the halo has any effect on cloud formation.

The bulge density profile is
\begin{equation}
\tilde{\rho}_b(r)=\rho_b\left(\frac{r}{R_e}\right)^{-p}{\rm e}^{-b(r/R_e)^{1/n}},
\end{equation}where $p=1-0.6097/n+0.05/n^2$ gives a S\'ersic profile with $n$ the S\'ersic index. $R_e$ is the radial scale parameter, and in GALACTICS $\rho_b$ is parametrized by the velocity parameter $\sigma_b\equiv\{4\pi n b^{n(p-2)}\Gamma[n(2-p)]R_e^2\rho_b \}^{1/2}$. We set these parameters to $n=1.31$, $\sigma_b=272$ km s$^{-1}$, and $R_e=0.64$ kpc.

We have named our fiducial run LowSoftMW. To test the effects of change 
the resolution, softening length, temperature floor, gas mass fraction, and artificial 
viscosity we investigate a total of ten different runs, summarized in Table~\ref{mwruntab}. Both MidSoftMW and HighSoftMW have higher gravitational 
softening lengths; MedGasMW and HighGasMW have higher gas mass fractions; LowFloorMW has a lower temperature floor; LowViscMW 
has lower artificial viscosity parameters ($\alpha$, $\beta$); and 
LowResMW has a lower resolution. In addition, as a convergence check we ran a higher resolution simulation (HighResFlatMW) with a total of $3.5\times 10^6$ particles and a softening length of $45$ pc, although we do not consider this our fiducial run as the simulation time did not reach a full Gyr. We found when running a simulation of this high resolution with identical initial conditions to LowSoftMW that the disc was dominated by a strong ring-shaped shock propagating outwards. This is caused by a combination of the rapid vertical collapse of the disc as it initially cools, and the rotation curve not being quite precise enough as GALACTICS is intended for collisionless mechanics and does not take into account gravitational softening or the pressure gradient of the gaseous disc. At the lower resolutions this shock is not well captured, and the disc quickly returns to equilibrium, so this is only a problem at our highest resolution. To prevent the shock becoming a problem it is necessary to start the simulation from an initially flattened state akin to the later evolution of the cooled disks. We therefore flattened the gas disc to a similar scale height as the cooled disks, which is a factor of 10 smaller. Circular velocities ($v_{\rm circ}$) were then set up using radial accelerations ($a_{\rm rad}$) generated from a single iteration of the HYDRA code, and explicitly setting $a_{\rm rad}=v_{\rm circ}^2/R$ for each gas particle, where $R$ is the radial coordinate of the particle. We also performed a simulation (FlatMW) with these initial conditions but at our fiducial (moderate) resolution, for a fair comparison of the effects of resolution. 

\begin{table*}
\begin{tabular}{ c | c | c | c | c | c | c | c | c | c |}
\hline\hline
Name & $l_{soft}$ (pc) & $T_{foor}$ (K) & $n_*$ & $n_{\rm g}$ & $n_{\rm DM} $ & $m_{\rm g}/m_{*}$ & $t_{\rm end}$ (Gyr) & $h_{\rm disc}$ (kpc) & $\alpha$, $\beta$ \\
\hline
LowSoftMW & $60$ & $300$&  $5\times10^{5}$ & $4\times10^{5}$ &  $5\times10^{5}$ & 0.1 & 1.116 & 0.36 & 1, 2\\
MedSoftMW & $100$ & $300$&  $5\times10^{5}$ & $4\times10^{5}$ &  $5\times10^{5}$ & 0.1 & 1.146 & 0.36 & 1, 2\\
HighSoftMW & $500$ & $300$&  $5\times10^{5}$ & $4\times10^{5}$ &  $5\times10^{5}$ & 0.1 & 1.542 & 0.36 & 1, 2\\
LowResMW & $60$ & $300$&  $1\times10^{5}$ & $8\times10^{4}$ &  $1\times10^{5}$ & 0.1 & 1.959 & 0.36 & 1, 2\\
LowFloorMW & $60$ & $10$&  $5\times10^{5}$ & $4\times10^{5}$ &  $5\times10^{5}$ & 0.1 & 1.004 & 0.36 & 1, 2\\
LowViscMW & $60$ & $300$&  $5\times10^{5}$ & $4\times10^{5}$ &  $5\times10^{5}$ & 0.1 & 1.002 & 0.36 & 0.5, 1\\
MedGasMW & $60$ & $300$ &  $5\times10^{5}$ & $4\times10^{5}$ &  $5\times10^{5}$ & 0.2 & 0.485 & 0.36 & 1, 2\\
HighGasMW & $60$ & $300$ &  $5\times10^{5}$ & $4\times10^{5}$ &  $5\times10^{5}$ & 0.5 & 0.434 & 0.36 & 1, 2\\
FlatMW & $60$ & $300$ & $5\times10^{5}$ & $4\times10^{5}$ &  $5\times10^{5}$ & 0.1 & 0.790 & 0.036 & 1, 2\\
HighResFlatMW & $45$ & $300$ &  $1.25\times10^{6}$ & $1\times10^{6}$ &  $1.25\times10^{6}$ & 0.1 & 0.318 & 0.036 & 1, 2\\
\hline
\end{tabular}
\caption{\label{mwruntab}Summary of Milky Way runs. $l_{soft}$ is the minimum softening length, $T_{floor}$ is the temperature floor, $n_*$, $n_{\rm g}$, and $n_{\rm DM}$ are the numbers of star, gas and dark matter particles, $m_{\rm g}/m_*$ is the gas/star mass ratio for the disc, $t_{end}$ is the total simulation time, $h_{\rm disc}$ is the scale height of the disc, and $\alpha$ and $\beta$ are the artificial viscosity parameters.} 
\end{table*}

\subsubsection{Monolithic Collapse Model}

This model consists of a spherically symmetric distribution of gas 
within an equilibrium NFW dark matter halo. We generate the halo using 
GALACTICS according to the parameters in Table~\ref{haloprams}, giving a 
halo with $M=1.1\times10^{12}M_\odot$.

For the gas we use the `high-entropy' profile of 
\cite{2009MNRAS.396..191K}, which was produced from equation 1 of 
\cite{2004ApJ...601...37K}, setting $c=1$, $\alpha=1$, $\beta=3$, and 
$\gamma=0$. Kaufmann noted that a gas density profile that is shallower 
than the NFW profile \citep[as expected in models with pre-heating feedback 
\eg][]{2002MNRAS.333..768M} produces an angular momentum 
distribution in the final object that better fits observations. In this 
model, the gas collapses into clumps which combine to form an unstable 
disc.

As in \citet{2007MNRAS.375...53K}, the initial temperature profile is 
calculated to provide hydrostatic 
equilibrium according to

\begin{equation}
T(r)=\frac{\mu}{k_B} \frac{1}{\rho_{\rm G}(r)} \int_r^\infty \rho_{\rm G}(r)\frac{GM_{\rm tot}(r)}{r^2} {\rm d}r,
\end{equation} 
where $\mu$ is the mean molecular 
weight of the gas (taken as its primordial value, $\mu\approx0.59 m_H$), 
$k_B$ is the Boltzmann constant, $\rho_{\rm G}$ is the initial gas density, 
and $M_{\rm tot}(r)$ is the total mass (gas and dark matter) within a sphere 
of radius $r$. We give the gas a flat velocity profile.
The positions of the gas particles in our initial conditions are simply 
the generated positions of the dark matter particles flipped as in section \ref{mwics}.

To set up a rotating halo, GALACTICS swaps a fraction of the dark matter 
particles' velocities over the radial axis to increase the number of particles rotating in the same direction. We assume the gas and the dark matter have the same specific rotational momentum, \ie
\begin{equation}
\frac{|L_{G}|}{M_{G}}=\frac{|L_{\rm DM}|}{M_{\rm DM}},
\end{equation} 
so that the spin parameter \citep{2008gady.book.....B} of the gas is equal to the spin parameter of the halo, 

\begin{equation}
\lambda_{G}=\frac{|L_{G}|}{M_{G}}\frac{\sqrt{|E_{\rm DM}|}}{GM_{\rm DM}^{3/2}}=\frac{|L_{\rm DM}|}{M_{\rm DM}}\frac{\sqrt{|E_{\rm DM}|}}{GM_{\rm DM}^{3/2}}=\lambda_{\rm DM}.
\end{equation}
We used a spin parameter of $\lambda_{G}=0.038$, close to the median value observed in simulations \citep{2001ApJ...555..240B,1987ApJ...319..575B}. After each gas/DM 
halo is produced, it is evolved for $0.5$ Gyr with cooling switched off 
to ensure the ICs are stable. Our first model (HighSoftC) is performed 
with the softening equal to \citet{2009MNRAS.396..191K}'s and the 
temperature floor equal to \citet{2009MNRAS.396..191K}'s cooling floor. 
We also investigated models with lower softening lengths and temperature 
floors to see if smaller clouds were resolved. A low resolution run was 
performed as a convergence check, and finally we performed a model with 
a low gas fraction to see the effect of increasing the disc's stability. 
These models are summarised in Table~\ref{cruntab}.

\begin{table*}
\begin{tabular}{ c | c | c | c | c | c | c |}
\hline\hline
Name & $l_{soft}$ (pc) & $T_{foor}$ (K) & $n_{\rm g}$ & $n_{\rm DM} $ & $m_{\rm g}/m_{\rm DM}$ & $t_{end} (Gyr)$\\
\hline
HighSoftC & $514$ & $3\times10^{4}$ & $5\times10^{5}$  &  $1\times10^{5}$ & 0.148 & 4.5\\
MidSoftC & $200$ & $3\times10^{4}$ & $5\times10^{5}$ &  $1\times10^{5}$ & 0.148 & 3.9\\
LowSoftC & $60$ & $3\times10^{4}$ & $5\times10^{5}$ &  $1\times10^{5}$ & 0.148 & 3.3\\
LowSoftFloorC & $60$ & $300$ & $5\times10^{5}$ &  $1\times10^{5}$ & 0.148 & 3.7\\
LowResC & $60$ & $300$ & $1\times10^{5}$ &  $1\times10^{5}$ & 0.148 & 4.6\\
LowMassC & $512$ & $3\times10^{4}$ & $5\times10^{5}$ &  $1\times10^{5}$ & 0.030 & 7.8\\
\hline
\end{tabular}
\caption{\label{cruntab}Summary of collapse runs. $l_{soft}$ is the minimum softening length, $T_{floor}$ is the temperature floor, $n_{\rm g}$ and $n_{\rm DM}$ are the numbers of gas and dark matter particles, $m_{g}/m_{\rm DM}$ is the gas/dark matter mass ratio, and $t_{end}$ is the total simulation time.} 
\end{table*}

\subsection{Cloud Identification \& Analysis}\label{energysection}
\subsubsection{Identification algorithm}

We identified clouds using a two-step process. A density 
threshold was applied and then particles above this threshold 
were linked into groups using the 
friend-of-friends algorithm. Using a density threshold partially avoids the notorious `string of pearls' effect that that may lead to spurious 
filamentary structures or the merging of many smaller structures into 
one larger one. We set the density threshold at a density
of $n=7$ $M_\odot {\rm pc}^{-3}$, and used a linking length of 50 
pc.

The high density threshold ensures that only dense cloud-like 
objects are selected, while the linking length is close to the size of 
the softening ensuring small 
fluctuations below this threshold can be skipped over. We set the 
minimum cloud size to $30$ particles, giving minimum cloud mass of 
$1.6\times10^5$--$8.5\times10^6 M_\odot$, depending on resolution. We 
found these parameters largely selected dense, cool clumps while 
excluding other objects such as filaments from cloud encounters. Because 
our lower limit for cloud size is a number of particles and not a mass, 
we include clouds of increasingly small mass as we increase the 
resolution of our simulation. This may create a resolution dependence 
until individual molecular clouds are resolved - a level that we do not 
achieve in this work, although as noted in section~\ref{stabilityanalysis} the major axisymmetric instabilities appear to be resolved.

Clouds are tracked from output to output by examining the particles 
resident in each cloud. If the cloud $A$
at time $t_i$ contains at least half of the particles contained by
cloud
$B$ at the time of the following output $t_{i+1}$, then $A$ is a parent of $B$. If $B$ contains at
least half of the particles contained by cloud $A$, then $B$ is a child
of $A$. If $B$ has several parents, then a merger has occurred. If $A$ has several children, then a separation has occurred. If
$A$ is the only parent of $B$, and $B$ is the only child of $A$, then
$B$ is identified as the {\em same cloud} as $A$.
This categorization allows for multiple parents to join in a merger and 
it is 
also possible 
for a parent to split into into multiple children. 
During simulations we observed that 
mergers can be complex with clouds merging and separating several times 
before settling into a single cloud, or in some cases while no longer 
interacting. This means our statistics are perhaps better thought of as 
recording `interaction' rates (including `self-interaction') rather than cloud collision rates.

\subsubsection{Treatment of cloud energy and interactions}
To quantify 
the energy loss due to interactions, we compare the kinetic and 
potential energies of clouds in sequential outputs across a separation 
or merger event. For the 
`combined' stage of the interaction (the earlier output for a 
separation, 
the later output for a merger) we calculate the centre of mass kinetic energy, 
\begin{equation}
{
K_{\rm combined}=\frac{1}{2}\left(\displaystyle\sum_{i}^{P}m_i\right)\left(\frac{\displaystyle\sum_{i}^{P}m_i\mathbf{v_i}}{\displaystyle\sum_{i}^{P}m_i}\right)^2
}
\end{equation}
where $P$ is the set of all particles `involved' in the interaction (defined below), and 
$\mathbf{v_i}$ and $m_i$ are the velocity and mass of particle $i$. This is 
compared with the sum of the centre of mass kinetic energies of the 
clouds during the `separated' stage of the interaction,

\begin{equation}
{
K_{\rm separated}=\displaystyle\sum_{j}^{C}\left\{\frac{1}{2}\left(\displaystyle\sum_{i}^{p_j}m_i\right)\left(\frac{\displaystyle\sum_{i}^{p_j}m_i\mathbf{v_i}}{\displaystyle\sum_{i}^{p_j}m_i}\right)^2\right\}
}
\end{equation}
where $p_j$ is the set of all particles in cloud $j$ and $C$ is the set 
of all clouds involved in the interaction during the separated stage.

The merged or unseparated cloud does not typically contain all of the 
particles from the clouds that formed it or that separated from it; 
there are always a number of particles that are expelled during the 
interaction, while other particles may accrete on to the clouds during 
the interaction. These particles carry kinetic energy, and so to ensure 
that we are measuring a real loss of energy from the system and not just 
an apparent energy loss from particles leaving the cloud phase, we define 
$P$ by 
\begin{equation}
{
P = \displaystyle\bigcup_j^C p_j.
}
\end{equation}

The final step in the energy budget calculation is to ensure that the 
energy change due to clouds moving in the potential well of the dark 
matter and baryons is accounted for. We take out the effects of these 
gravitational interactions by calculating the potential between $P$ and 
all other particles during both dumps, and subtracting the difference from the 
kinetic energy that was calculated.

\subsubsection{Energy analysis}
From the cloud energy budget we can obtain an estimate for the total 
time-scale 
for dissipation of kinetic energy from cloud-cloud interactions. By 
analogy with the star-formation time-scale, typically defined as 
$t_{SFR}=\Sigma_{gas}/(d\Sigma_{*}/dt)$, we define the viscous time-scale 
due to cloud-cloud collisions to be the dissipative time-scale,

\begin{equation}
{
t_{\nu_{\rm col}}=\frac{K}{-{\rm d}K_{C}/{\rm d}t},
}
\end{equation}
where $K$ is the total rotational kinetic energy of the gas, and $-{\rm 
d}K_{C}/{\rm d}t$ is the rate at which this kinetic energy is dissipated 
due to collisions. This dissipation rate can be written as

\begin{equation}
{
\frac{-{\rm d}K_{C}}{{\rm d}t}=C \frac{\Delta K_{\rm col}}{K},
}
\end{equation}
where $C$ is the interaction rate (determined by counting the 
number of interactions that occur within a time period) and $\Delta 
K_{\rm col}$ is the energy lost per interaction. In practice, we average over $n_{\rm col}$ interactions so that

\begin{equation}
{
t_{\nu_{\rm col}} = \frac{\Delta t}{n_{\rm col}} \frac{\sum_{i=1}^{n_{\rm col}} K(t_i)}{\sum_{i=1}^{n_{\rm col}}\Delta K_i}.
}
\end{equation}
where $\Delta t$ is the time period that the $n_{\rm col}$ interactions 
occurred over (and hence $C=\frac{n_{\rm col}}{\Delta t}$), $\Delta K_{i}$ 
is the kinetic energy lost in a particular interaction $i$ and $K(t_i)$ 
is the total kinetic energy in gas at the time of that interaction.

It is important that we connect this method of measuring the dissipative 
time-scale in our models with definitions used elsewhere. It is commonly 
argued \citep[\eg][]{2002ApJ...581.1013B} that the form of the
viscous time-scale is
\begin{equation}
t_{\nu}\approx \frac{R^2}{\nu},
\end{equation} where $R$ is the radial coordinate and $\nu$ the (effective) viscosity.

To see how this form arises in our measurements, consider the 
following argument: If we neglect radial velocity, then the kinetic 
energy per unit volume of a component of fluid in a rotating disc is 
$k=\rho (R\Omega)^2/2$. We can convert the rate of viscous dissipation for a generic fluid ($\Phi$, the energy lost per unit volume per unit time) from \citet{1984frh..book.....M} 
into cylindrical coordinates and again assume angular velocity 
dominates, simplifying it to:
\begin{equation}
\Phi=\rho\nu (R\Omega^\prime)^2,
\end{equation} where the prime indicates a radial derivative. We can substitute these values into our definition for $t_{\nu_{\rm col}}$ 
because $\Phi=-{\rm d}k_{C}/{\rm d}t$, so
\begin{equation}
t_{\nu_{\rm col}} = \frac{\rho (R\Omega)^2/2}{\rho \nu (R\Omega^\prime)^2} = \frac{\Omega^2}{2\nu(\Omega^\prime)^2}.
\end{equation}
If we then take a power law for rotation $\Omega\propto R^{-\alpha}$ then
\begin{equation}
t_{\nu_{\rm col}} = \frac{1}{2\alpha^2}\frac{R^2}{\nu},
\end{equation} 
which agrees with $R^2/\nu$ within a factor of $1/2\alpha^2$. For a flat 
rotation curve, $\alpha=1$ and this factor is merely $1/2$ - hence the 
dissipative time-scale is of the order of the traditional viscous 
time-scale. Note, \citet{1987ApJ...320L..87L} give a different prefactor 
- 
$(2-\alpha)/(\alpha)$. However, these values all agree within an order 
of magnitude, provided $\alpha$ is not extremely large or small. 
Although our viscous time-scales are calculated over the whole disc to 
ensure sufficient numbers of interactions are measured, and the 
analytical $R^2/\nu$ is a local value at a specific radius, we should not expect this to have 
an effect beyond an order of magnitude, assuming analytical viscous 
time-scales have been calculated at a representative radius.

\section{Results}\label{results}

\subsection{Milky Way Model}

\subsubsection{General Evolution}

The evolution of all models excluding HighSoftMW are 
similar\footnote[1]{Animations for some models presented here are 
available at\\{\url 
{http://ap.smu.ca/~thacker/williams/cloudcols.html}}}. In these models 
the gas disc is initially close to equilibrium. However, the gas rapidly 
cools and becomes unstable, collapsing vertically (except in FlatMW and HighResFlatMW, which are produced from already-collapsed initial conditions), and forming spiral instabilities which fragment into large number 
of small ($m\sim10^6$--$10^7 M_\odot$, $R\sim100$ pc) clouds.

\begin{figure*}
\begin{center}
\includegraphics[width=1.\columnwidth]{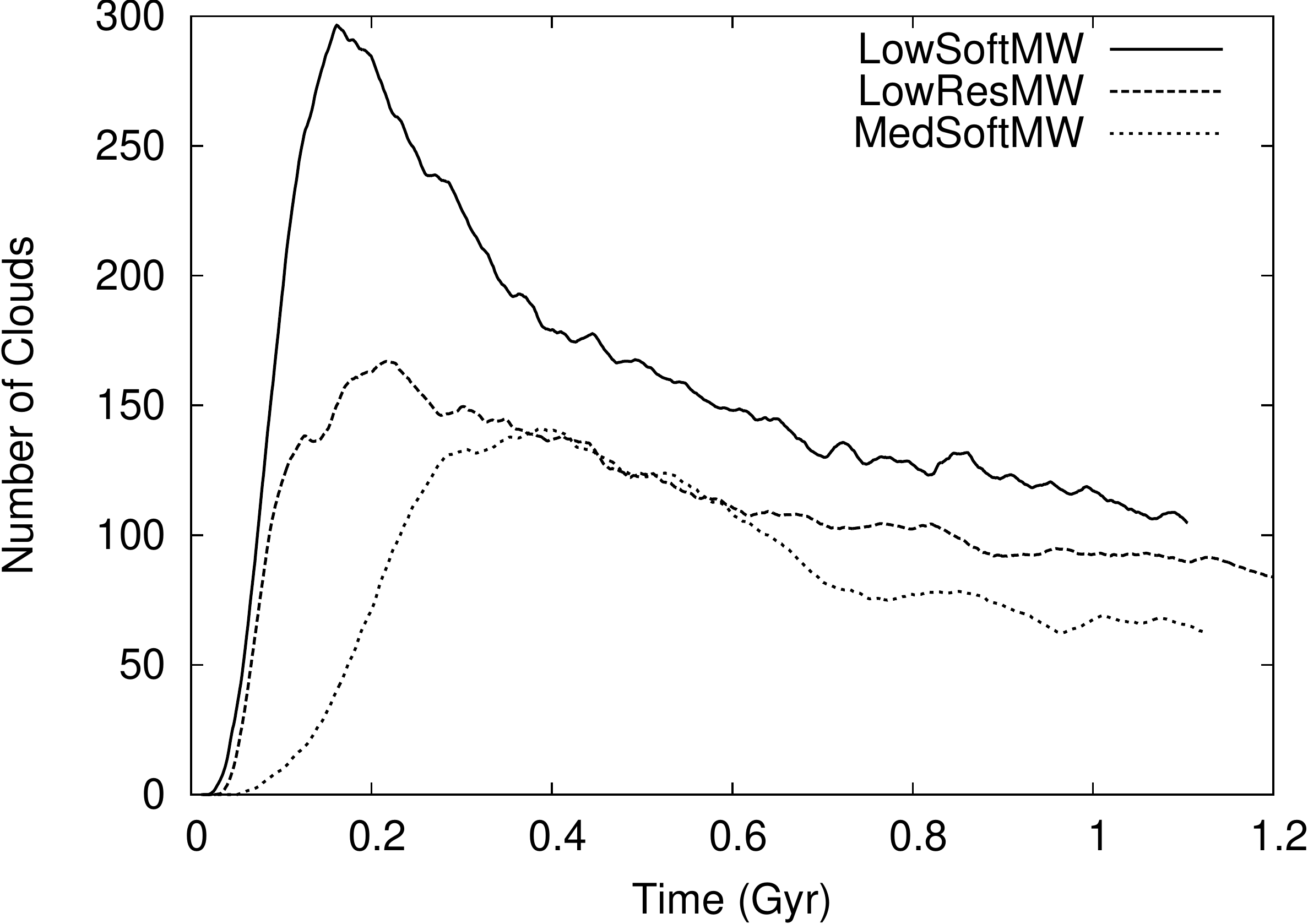}~\includegraphics[width=1.\columnwidth]{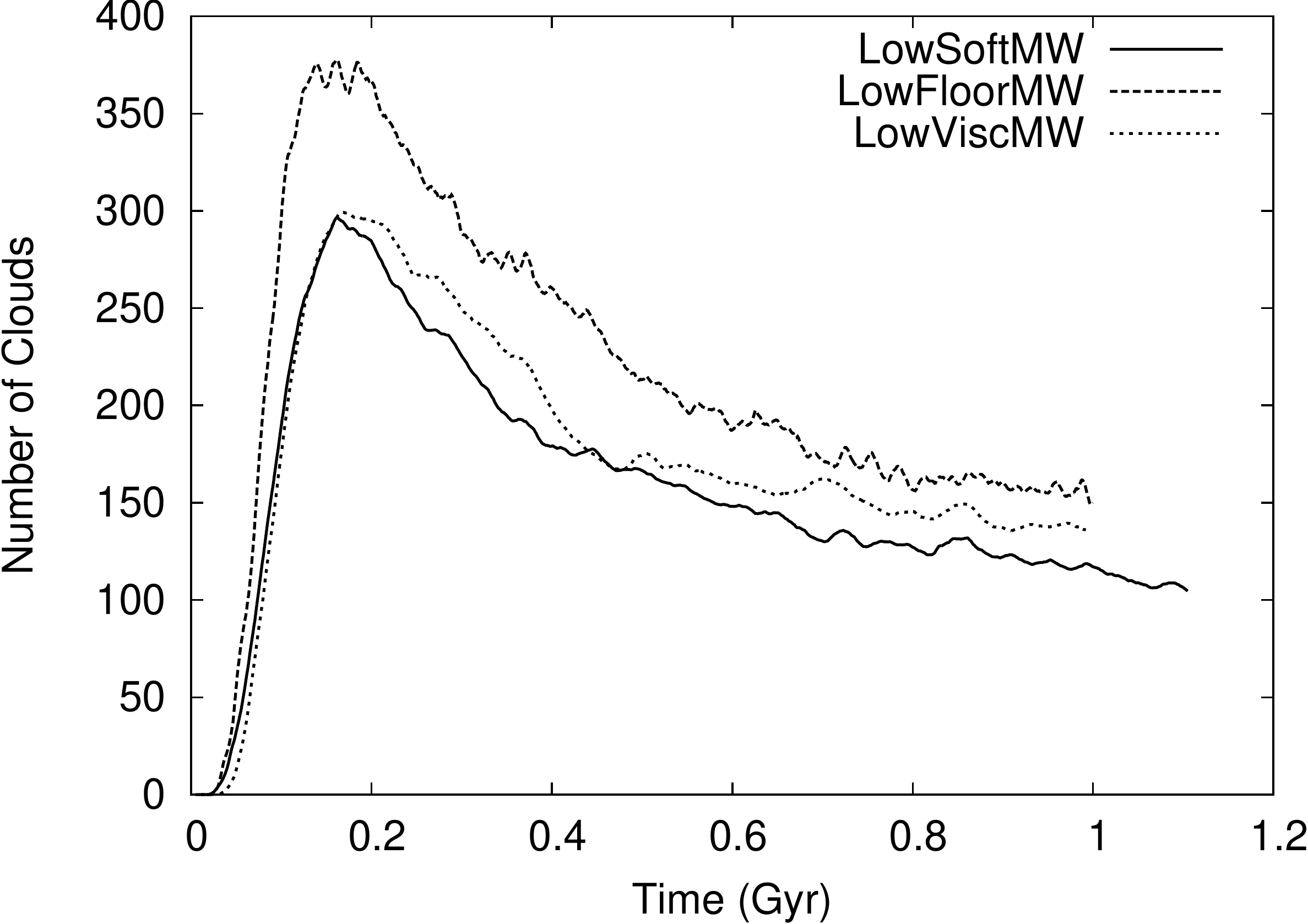} \\
\includegraphics[width=1.\columnwidth]{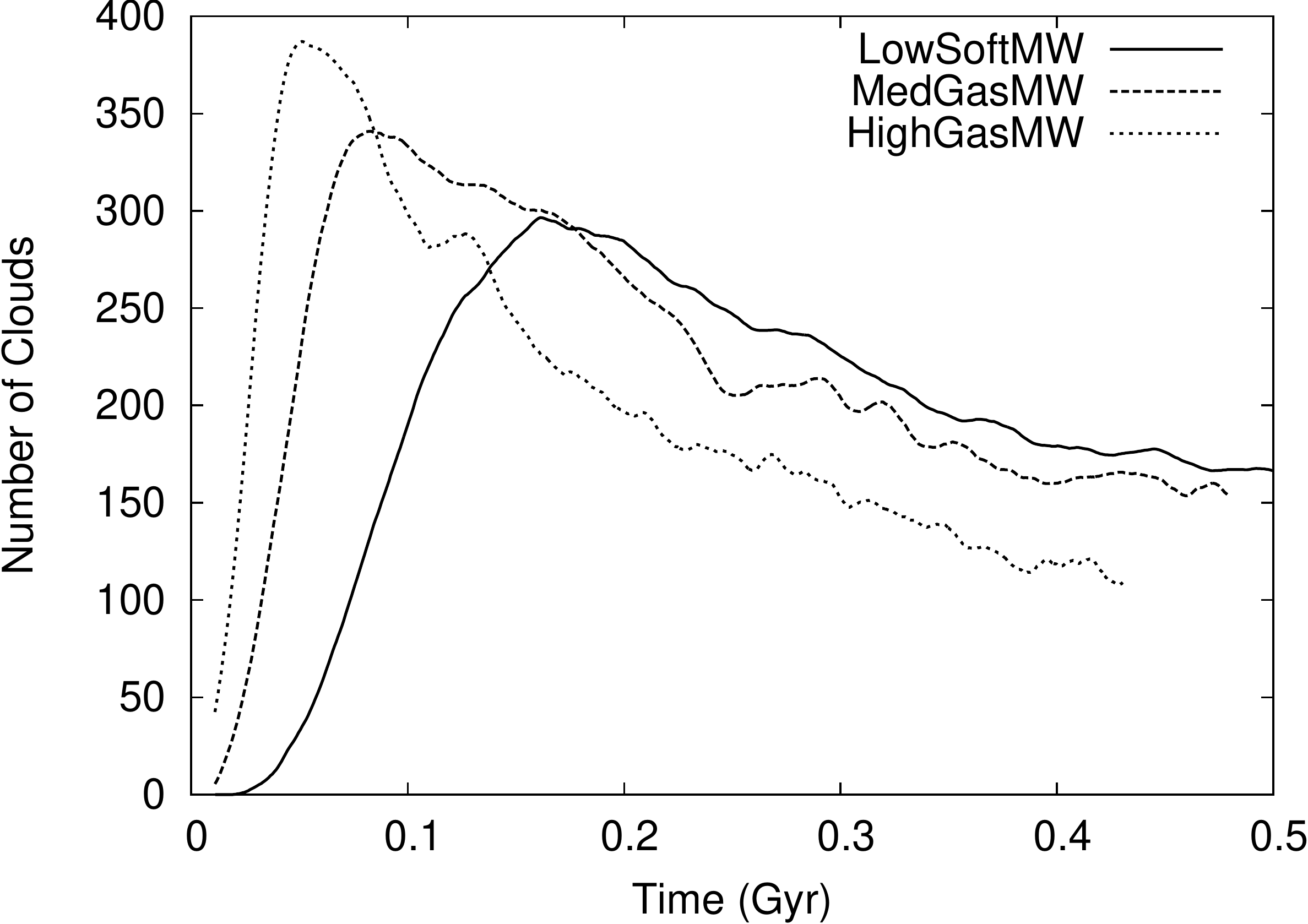}~\includegraphics[width=1.\columnwidth]{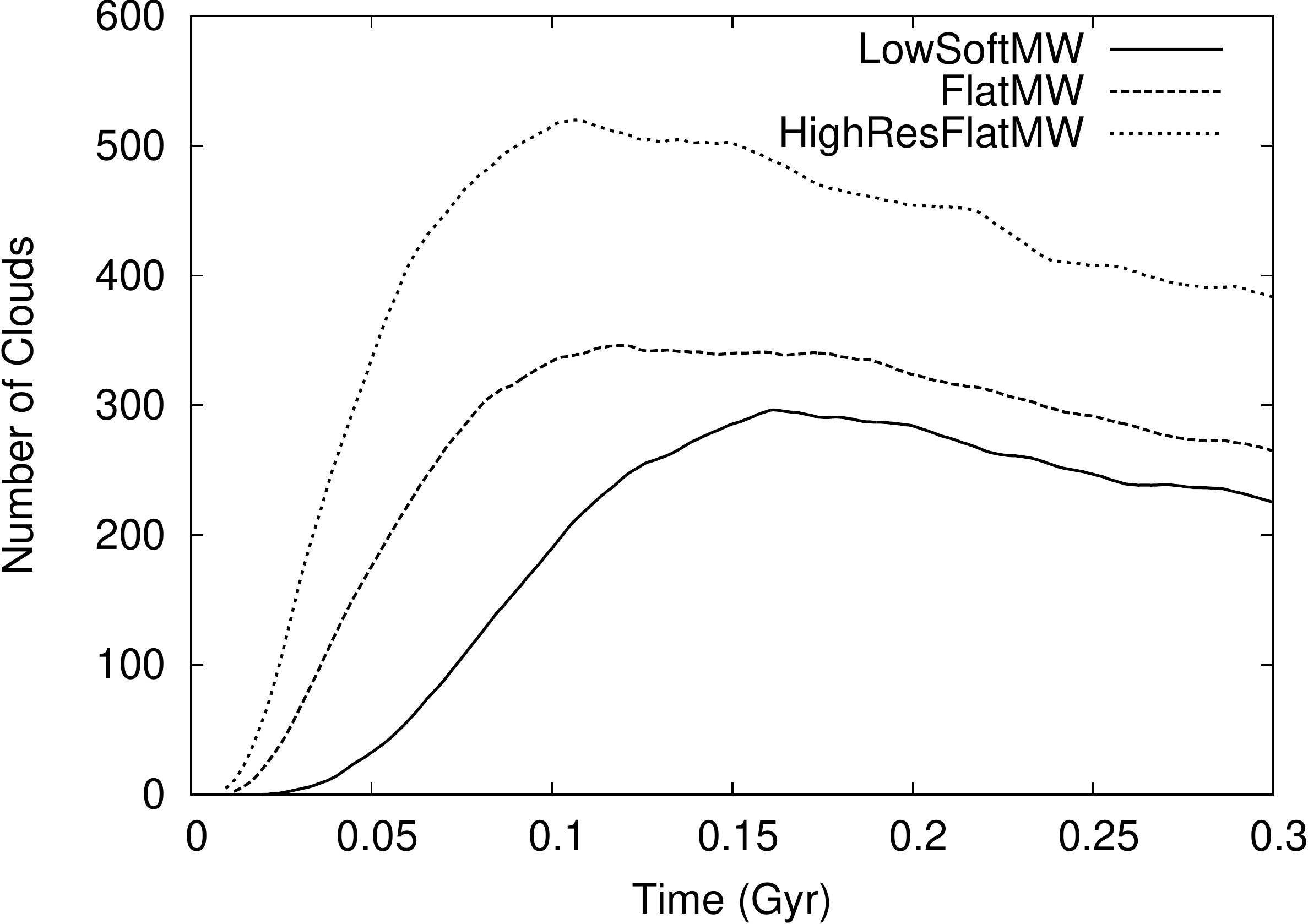} 
\end{center}
\caption{\label{ncMW} 	
Number of clouds in Milky Way models. To smooth the data, each plotted point is an average of the $29$ data points centred on it. The number peaks when many clouds are rapidly formed as the gas temperature drops below the Toomre instability 
threshold. It drops as these clouds merge.
}
\end{figure*}

\begin{figure*}
\begin{center}
\includegraphics[width=1.\columnwidth]{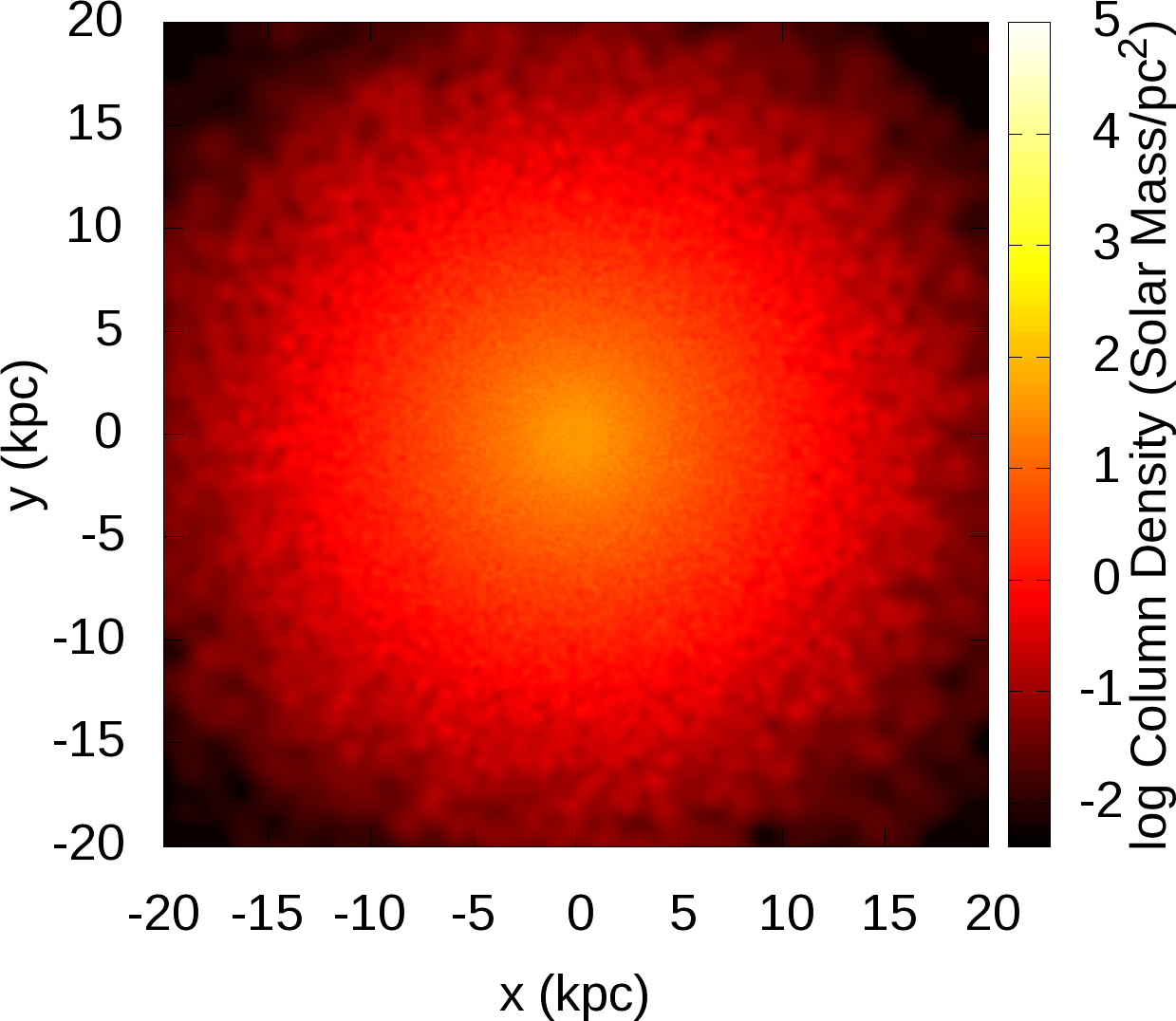}~\includegraphics[width=1.\columnwidth]{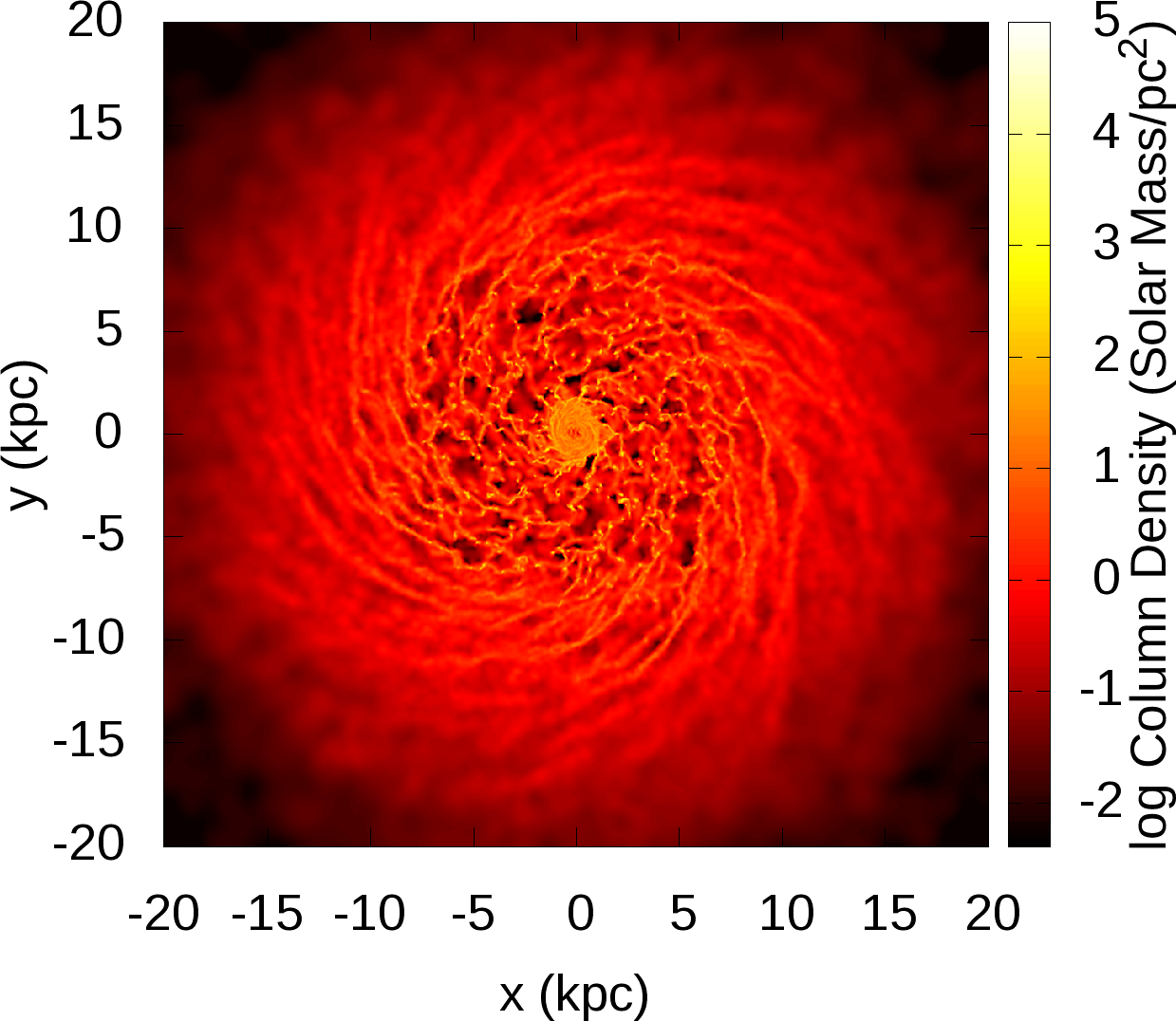}\\
~\\
\includegraphics[width=1.\columnwidth]{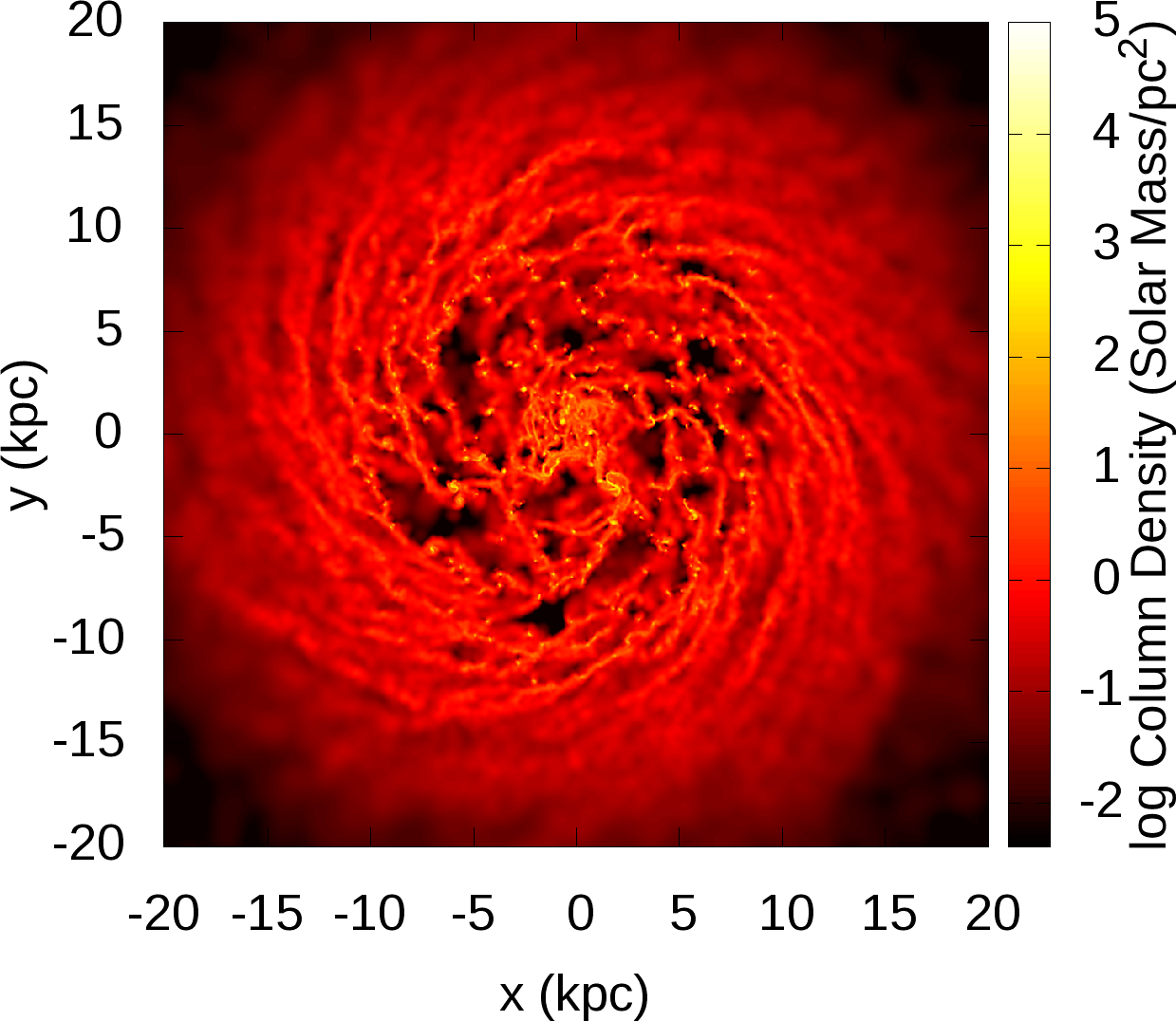}~\includegraphics[width=1.\columnwidth]{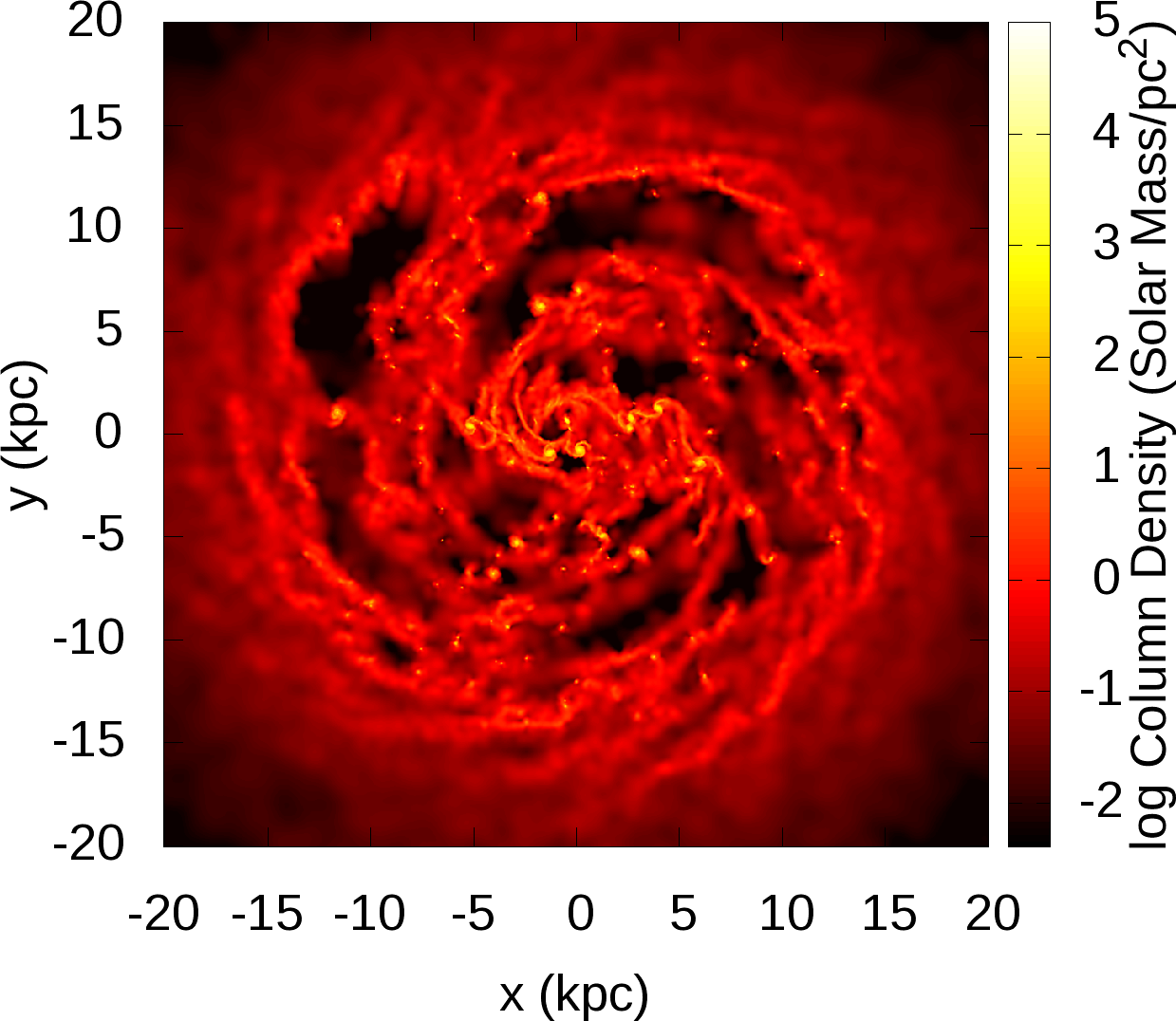}\\
\end{center}
\caption{\label{faceonlowsoftmw} 
Evolution of LowSoftMW. A featureless disc (top-left) rapidly collapses 
into a larger number of clouds (top-right) after around $200$ Myr of 
evolution. These clouds interact with each other and accrete material from $400$ Myr (bottom-left) until the simulation 
ends after $1.1$ Gyr (bottom-right). }
\end{figure*}

\begin{figure}
\begin{center}
\includegraphics[width=1.\columnwidth]{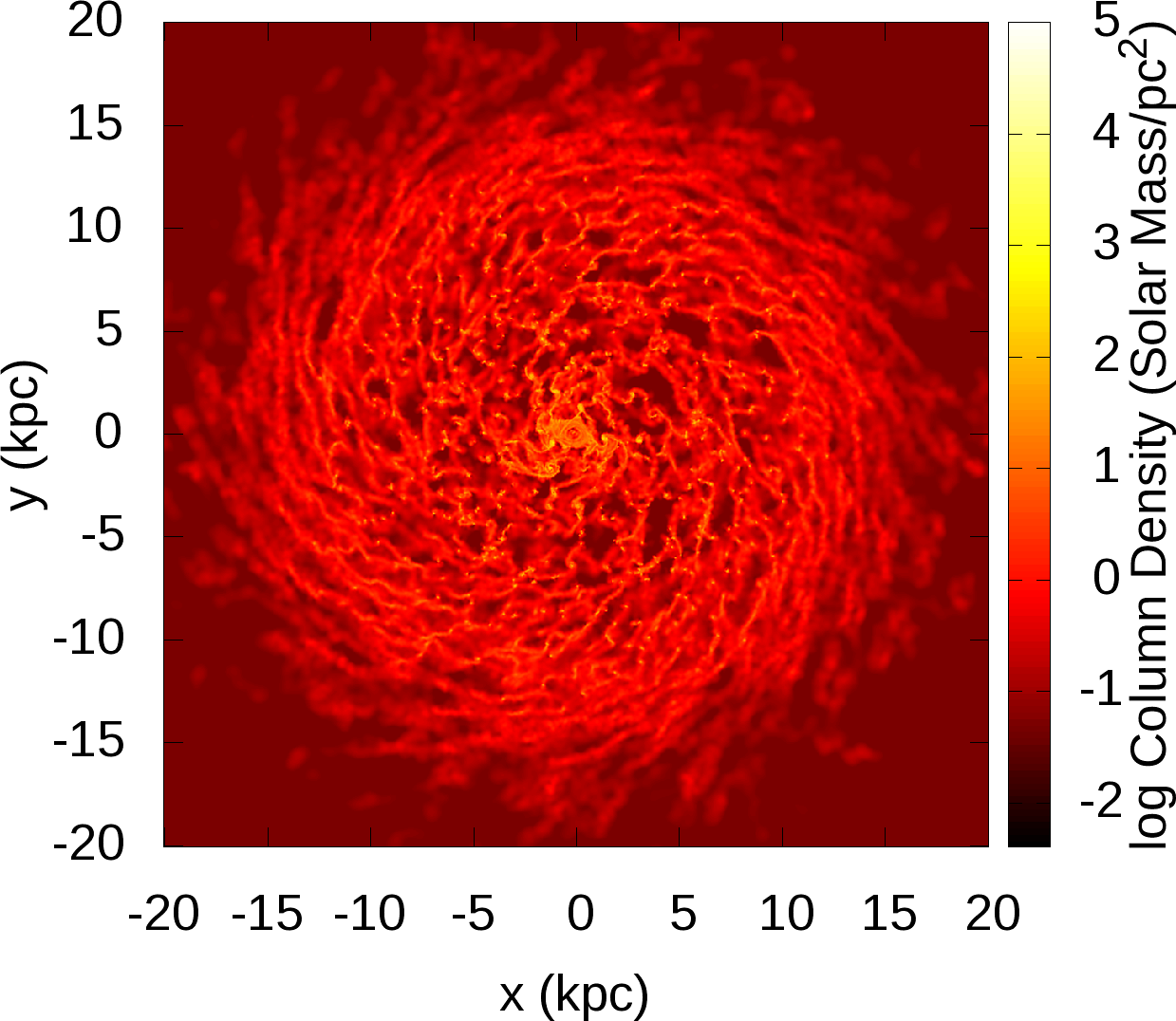}
\end{center}
\caption{\label{faceonhighresmw} 
HighResFlatMW after $300$ Myr of evolution}
\end{figure}

\begin{figure}
\begin{center}
\includegraphics[width=1.\columnwidth]{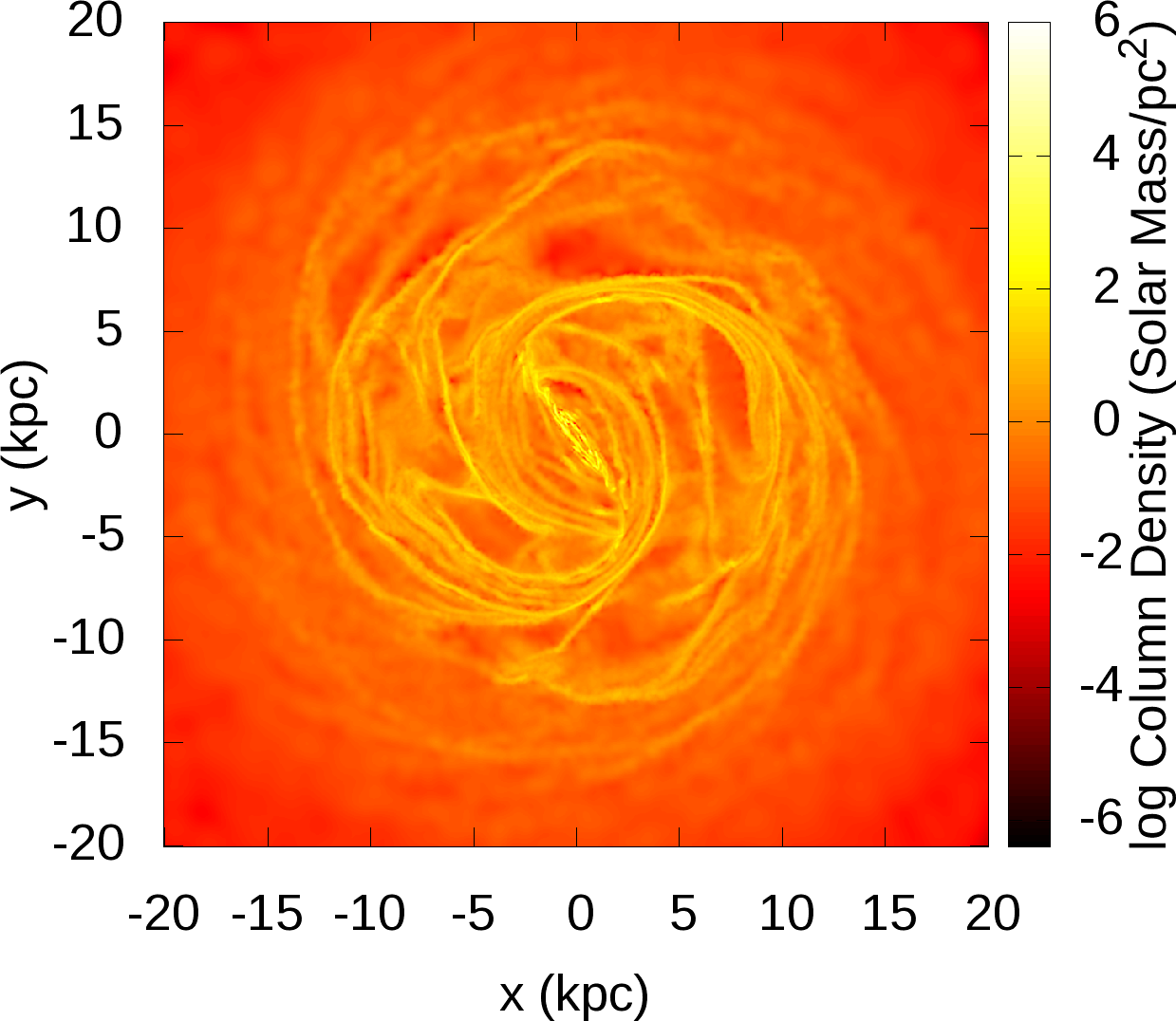}
\end{center}
\caption{\label{faceonhighsoftmw} 
HighSoftMW after $\sim1.5$ Gyr of evolution. Because of the large softening 
length, the disc does not 
undergo local fragmentation 
into clouds, and is instead dominated by bar and spiral 
instabilities.}
\end{figure}

After this 
epoch of rapid cloud formation, the clouds merge and continue to 
accrete material. The number of clouds drops, as illustrated in 
Fig.~\ref{ncMW}, while the total mass within clouds continues to 
increase until both reach a less dramatic stage from around $0.8$--$1.0$ Gyr, where the number of clouds decays only gradually as the mass within clouds gradually increases. A face-on view of the evolution of LowSoftMW is 
shown in Fig.~\ref{faceonlowsoftmw}, and a snapshot of HighResFlatMW is shown in Fig.~\ref{faceonhighresmw}. In HighSoftMW cloud collapse was 
quenched by the high softening length, and instead the disc was 
dominated by large scale instabilities (Fig.~\ref{faceonhighsoftmw}). The higher gas mass in HighGasMW and MedGasMW reduced the hydrodynamic time-step and so these simulations could only be run for $\sim0.45$ Gyr, while the increased computational load of the high resolution run HighResFlatMW also made a full simulation of $1.0$ Gyr unfeasible, and so this simulation was evolved for $\sim0.3$ Gyr.

The gas disc separates into two phases: diffuse 
gas which retains a moderate temperature ($\sim10^3$ to $\sim10^4$ K) 
though shock 
heating and a low 
cooling time, and dense gas whose temperature is tightly controlled by 
the Robertson-Kravstov dynamic temperature floor. It should be noted 
that while our models lack direct stellar feedback, the dynamic floor 
can heat the dense gas to temperatures as high as $3\times10^4$K. This 
temperature is equivalent to a sound speed of $\sim26$ km s$^{-1}$, which is on 
the order of the velocity dispersion generated by various feedback 
mechanisms \citep{2000ApJ...545..728T,2007MNRAS.374.1479G,2011ApJ...731...41O}. Hence 
while we expect 
implementing feedback would change our results, the difference may not 
be large. This is further supported by the findings of 
\citet{2008ApJ...684..978S}
, who found that the properties of large 
clouds are not strongly sensitive to feedback. Tests were also performed 
with a higher cooling floor of $3\times10^4$ K, and no clouds were 
formed. This perhaps demonstrates that a static cooling floor is a worse 
approximation to feedback as it inputs energy into any cool region of 
gas regardless of density, impeding any collapse that would have 
actually formed stars, in contrast with a dynamic temperature floor 
which 
inputs energy only into dense star-forming gas.

\subsubsection{Cloud formation \& numerical issues}\label{stabilityanalysis}

We now draw attention to the differences between the simulations illustrated in Fig.~\ref{ncMW}. While LowResMW produces clouds at the same time as LowSoftMW (top left), it produces fewer of them as the mass spectrum is truncated. Simlarly, FlatMW produces clouds at the same time as HighResFlatMW, but in smaller numbers (bottom right). Hence there is a trend of producing more clouds with increasing resolution. Overall, the flat initial conditions of FlatMW and HighResFlatMW produced clouds earlier and in greater numbers than in LowSoftMW. LowViscMW appears identical to LowSoftMW, suggesting that numerical artefacts due to artificial viscosity are not a significant effect (top right). LowFloorMW produced more clouds than LowSoftMW as the lower cooling floor allows the disc to become more unstable to cloud formation from Toomre instabilities. We also found that clouds formed earlier and were more numerous with increasing gas fraction, as demonstrated by HighGasMW and MedGasMW (bottom left).

\begin{figure*}
\begin{center}
\includegraphics[width=1.\columnwidth]{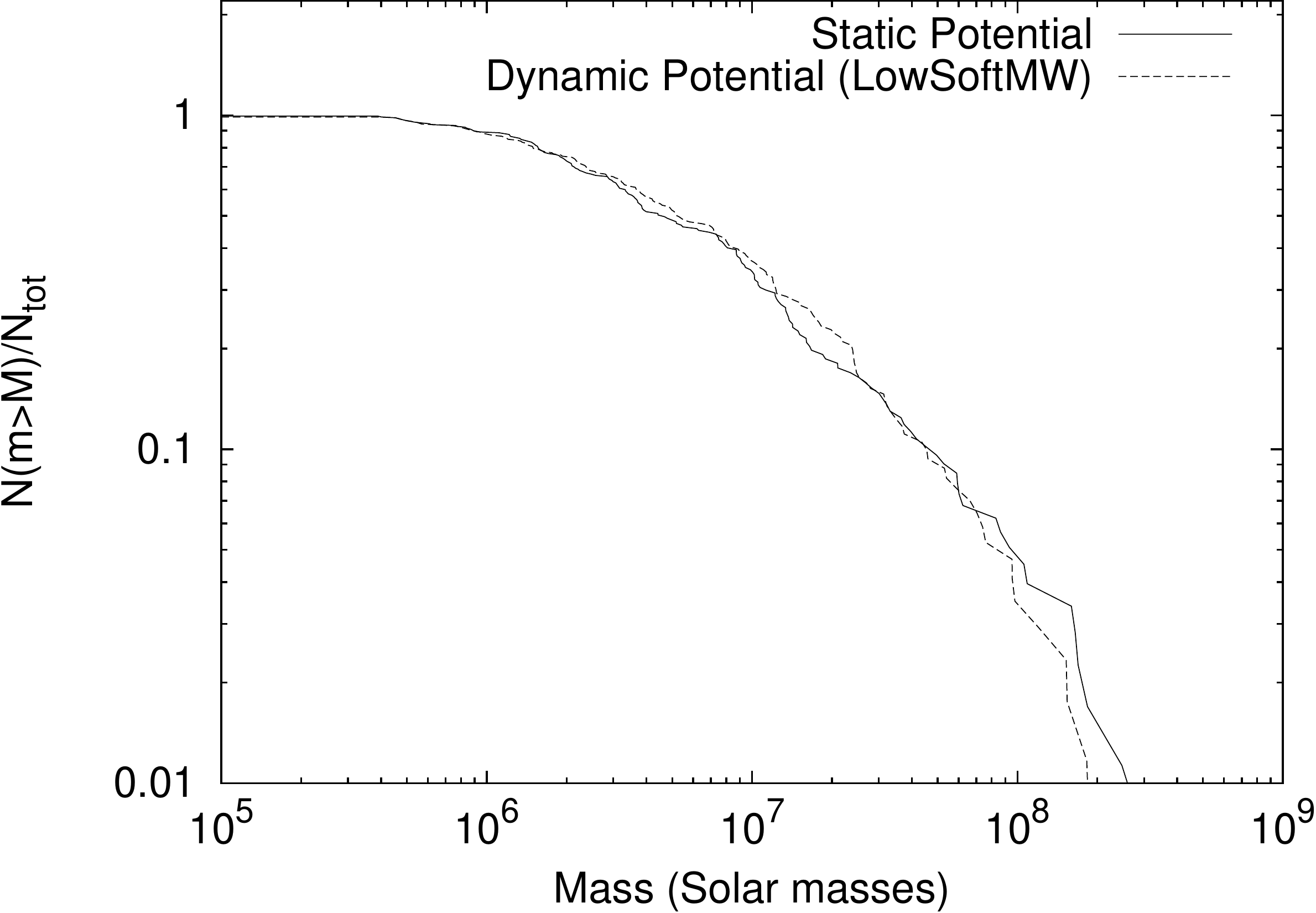}~
\includegraphics[width=1.\columnwidth]{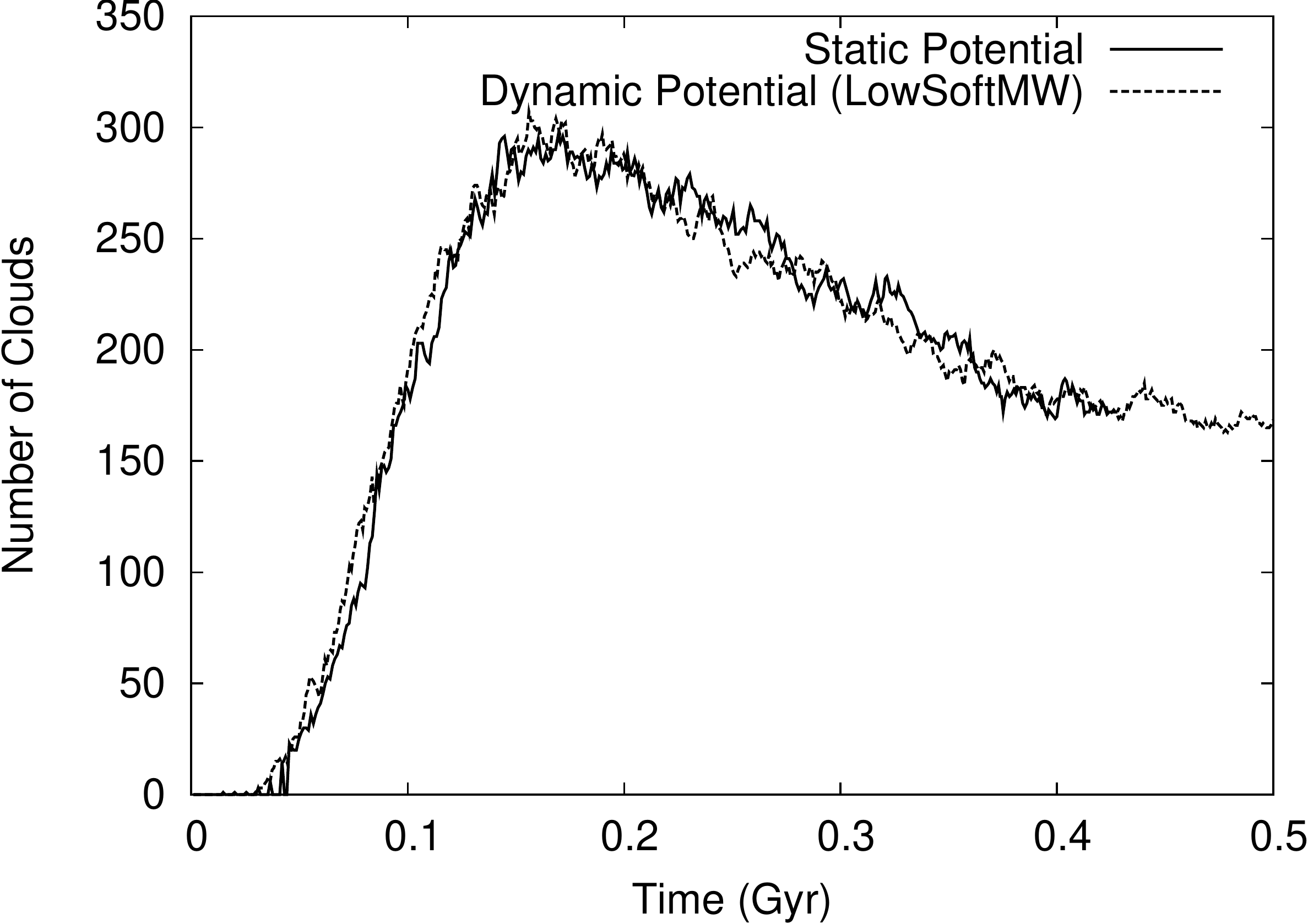}\\
\end{center}
\caption{\label{staticpot} 
Mass spectra at $430$ Myr (top) and cloud counts (bottom) for the fiducial model (LowSoftMW) and a test run with a static analytic potential.}
\end{figure*}

We found that replacing the halo with a static potential did nmot have a significant effect - the mass spectra and number of clouds formed over $430$ Myr of evolution were almost identical (Fig.~\ref{staticpot}).

As expected, the gravitational softening parameter has a 
significant 
effect on cloud formation. With a softening of $60$ pc 
(LowSoftMW), a maximum of $\sim300$ clouds were formed at a time of 
$0.02$ Myr, while with a softening of $200$ pc (MidSoftMW), half as many 
were formed 
($\sim150$), and the peak number was achieved later ($0.04$Myr). It 
should be noted though, that 
both 
models 
have a similar fraction of mass in clouds ($\sim80\%$). Increasing the 
softening yet further to $500$ pc (HighSoftMW), leads to almost no clouds 
forming other 
than a few clouds in the centre of the galaxy after about a Gyr of 
evolution (not shown in Fig.~\ref{ncMW}). These results match what would be expected on theoretical 
grounds.
Increasing the softening length delays cloud formation and 
produces fewer, more massive clouds, unless the softening length is 
increased above a certain threshold, beyond which cloud formation is 
prevented. 

It seems most likely that this threshold softening 
length is related to the wavelength of the unstable mode that causes 
cloud formation. We can calculate this using the two-fluid (gas/star) 
$Q_{gs}$ stability parameter from 
\citet{1984ApJ...276..114J}, \citet{2001MNRAS.323..445R}, and \citet{2005ApJ...626..823L}.
The 
individual Q parameters for stars and gas are defined as

\begin{equation}
Q_s=\frac{\kappa \sigma_s}{\pi G \Sigma_s}, Q_{\rm g}=\frac{\kappa c_{\rm g}}{\pi G \Sigma_{\rm g}},
\end{equation} 
where $\Sigma_s$ and $\Sigma_{\rm g}$ are the gas and stellar surface 
densities, $\sigma_s$ the stellar radial velocity dispersion, $c_{\rm g}$ the 
gas sound speed, and $\kappa$ the epicyclic parameter. Note that $Q_s$ 
differs from \citet{1964ApJ...139.1217T}'s definition of $Q$ for a 
collisionless system by a factor of $3.36/\pi$. If we define

\begin{equation}q=2\pi\sigma_s/(\kappa\lambda_i), f=c_{\rm g}/\sigma_s, \end{equation}where $\lambda_i$ is the wavelength of a particular mode of instability, and treat the stars as a fluid with sound speed equal to $\sigma_s$ as in \citet{2001MNRAS.323..445R} \cite[who follows][]{1984ApJ...276..114J}, we can define a combined $Q_{gs}$ by
\begin{equation}\frac{1}{Q_{gs}}=\frac{2}{Q_s}\frac{q}{1+q^2}+\frac{2}{Q_{\rm g}}\frac{fq}{1+q^2f^2},\end{equation} with a stability condition of $Q_{gs}<1$.

\begin{figure}
\begin{center}
\includegraphics[width=1.\columnwidth]{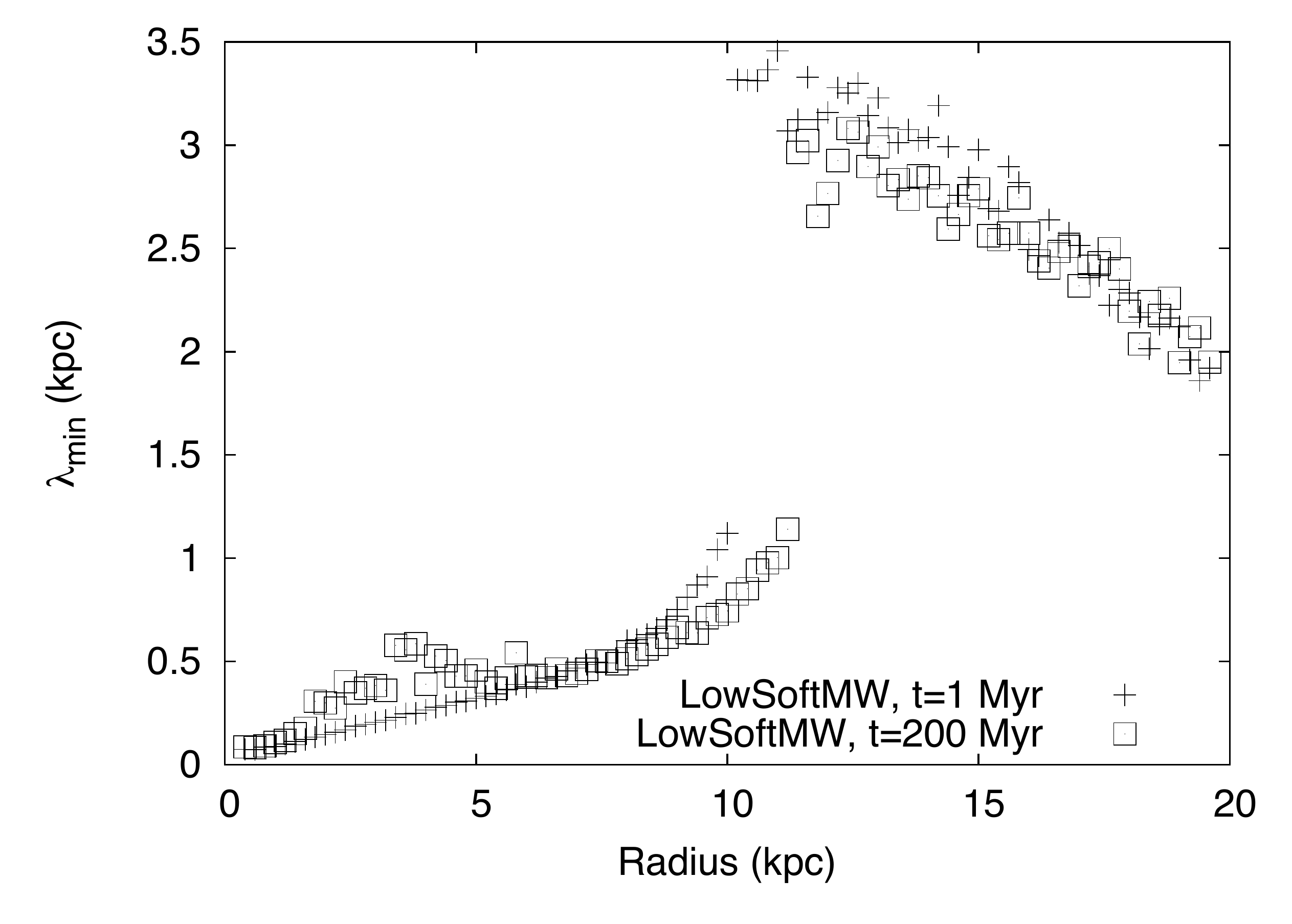}
\end{center}
\caption{\label{unstableplot} Wavelength $\lambda_i$ of the most unstable mode for LowSoftMW at $1$ and $200$ Myr}
\end{figure}

We calculate $Q_{gs}$ by using azimuthal means of $\Omega$, $\Sigma$, 
$\kappa$, $c_{\rm g}$ and $\sigma_s$, and setting $\lambda_i$ to $\lambda_{min}$, the wavelength 
that minimizes $Q_{gs}$. It is worth cautioning that these parameters 
are derived from linear 
perturbation theory and may not adequately describe the system once 
clouds have formed. Nevertheless, $\lambda_{min}$ does not rapidly vary 
from $t=1$ Myr to $t=200$ Myr for LowSoftC as shown in 
Fig.~\ref{unstableplot}. $\lambda_{min}$ is fairly small ($<1 kpc$) 
until a radius of 
$10$ kpc at which point it triples in size. This jump is due to the small wavelength gas instabilities starting to dominate over the large wavelength stellar instabilities.
A comparison with the face-on density plots (e.g. 
Fig.~\ref{faceonlowsoftmw}) shows that clouds form within $10$ kpc. 
In this region $\lambda_{min}$ is of the order of $100$s of pc. 
The `threshold' resolution for cloud formation (assuming 4 to 5 
softening lengths are required) in our models lies 
somewhere between $200$ pc and $500$ pc, and is consistent with this 
range. This quantifies an often quoted caveat for galaxy models - if the 
gravitational softening length is larger than the wavelength of the most 
unstable modes, then fragmentation is artificially frustrated.

The size of the unstable perturbations can be used to crudely estimate 
the masses of clouds. Assuming that the disc fragments into clumps 
of mass $\sim\pi\Sigma 
\lambda_{min}^2$, then for the LowSoftMW simulation (for example) the typical 
cloud masses should be 
the order of several $10^6 M_{\odot}$, which is admittedly 
significantly larger than average molecular cloud masses and actually 
much 
closer 
to giant molecular cloud complex masses. Nonetheless, this value 
is broadly consistent with 
our 
spectrum of cloud masses (e.g. Fig.~\ref{mwms}). However, we caution against 
over interpretation as the mass spectrum convolves together an initial 
spectrum and its subsequent evolution. If this simple approach to 
calculating initial cloud masses were accurate we
would not expect a higher resolution model to produce smaller clouds
from this mode of instability, although non-azimuthally symmetric modes which may produce smaller scale instabilities have been excluded from this analysis.
Smaller clouds could also 
be produced in
a higher resolution Milky Way model by changing the initial conditions,
or if these giant clouds undergo further fragmentation.

\subsubsection{Cloud Mass Functions}

\begin{figure*}
\includegraphics[width=1.0\columnwidth]{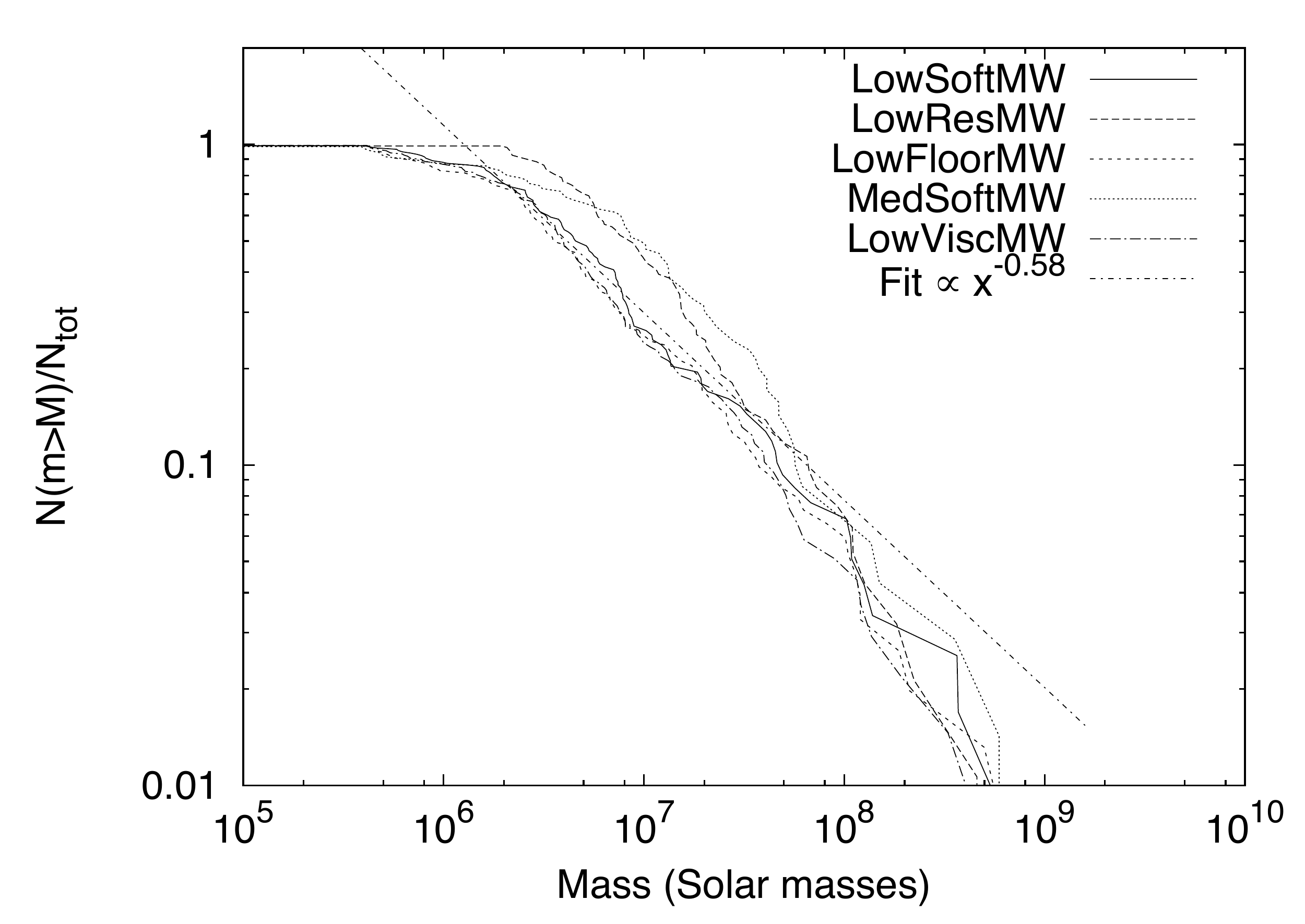}\includegraphics[width=1.0\columnwidth]{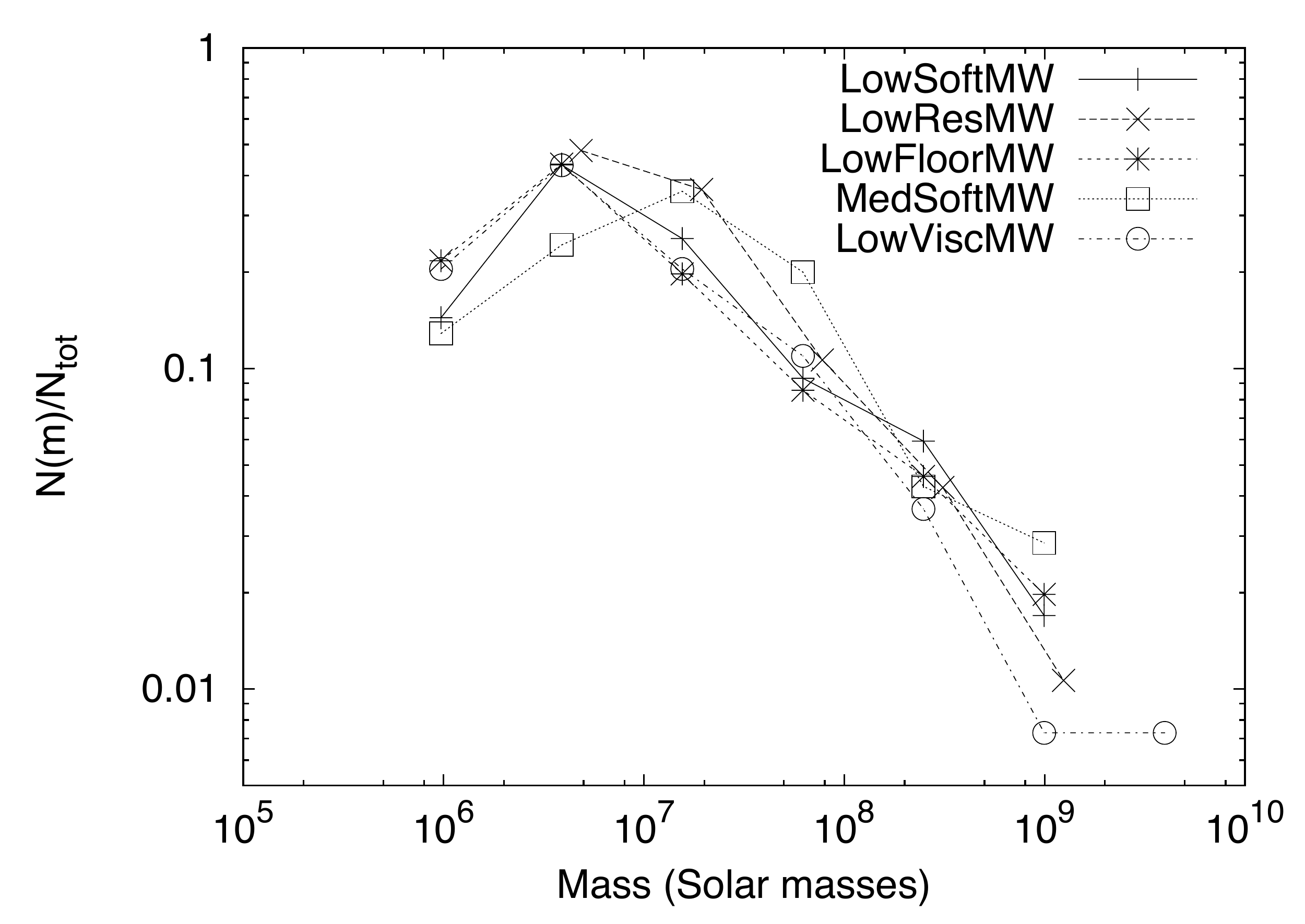}\\
\caption{\label{mwms} Mass spectra for clouds in Milky Way runs at $800$ Myr. Left: Cumulative mass spectra \citep[for comparison with][]{2009MNRAS.392..294A}. Right: Differential mass spectra (for comparison with \citet{taskertan}). The bins in the differential mass plot have a width of $\log (4)\approx0.6$ dex.}
\end{figure*}

\begin{figure}
\includegraphics[width=1.\columnwidth]{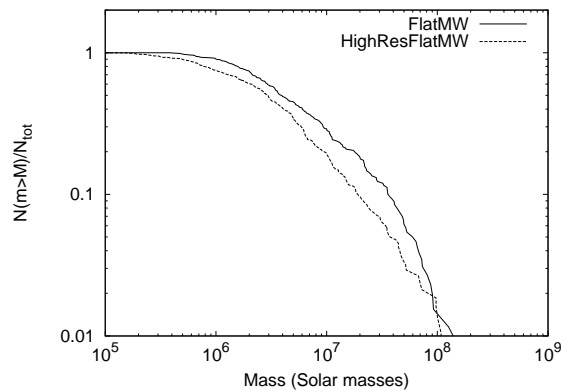}\\
\caption{\label{spectres} Cumulative cloud mass spectra from flat initial conditions, including our highest resolution model.}
\end{figure}

The mass functions of our clouds (Fig.~\ref{mwms}) differ from those of 
\citet{taskertan} and \citet{2009MNRAS.392..294A} in that our clouds are 
more 
massive. 
However, neither of these studies has equivalent physics.  Tasker \& 
Tan differ in that they do not include dynamic stellar disc while  
Agertz et al include feedback. Resolution could potentially 
also be an 
issue: although our mass function does not greatly vary between our low and moderate 
resolution models in our fiducial simulations, our high resolution flat model produced lower mass clouds than the moderate resolution flat model (Fig.~\ref{spectres}).

The high-mass region of our cumulative mass spectrum plot 
follows a power law (i.e. $N(m)\propto m^\alpha$ or $N(m>M)\propto 
M^{\alpha+1}$). A least-squares fitting gives $\alpha\sim-1.5$. This 
is slightly shallower than the $\sim-1.8$ in the simulations of 
\citet{1996ApJ...462..309D}, and \cite{2008MNRAS.385.1893D}
but close to the values of $-1.5$ to $-1.6$ from observations 
\citep{1985ApJ...289..373S,1987ApJ...319..730S,1989ApJ...339..919S,1997ApJ...476..166W,2010ApJ...723..492R}, and from the simulated mass spectra at around $10^6 M_\odot$ at $300$ Myr in \citet{taskertan} and at $1$ Gyr in \citet{2009MNRAS.392..294A}.

\subsubsection{Viscous time-scales}\label{mwviscsec}

\begin{table}
\begin{tabular}{ c | c | c | c |}
\hline\hline
~ & ~ & Viscous & Simulation\\
Name & Interactions & time-scale (Gyr) & time (Gyr)\\
\hline
HighSoftC & 104 & 2.0 & 4.5\\
MidSoftC & 211 & 1.8 & 3.9\\
LowSoftC & 566 & 0.8 & 3.5\\
LowSoftFloorC & 3672 & 5.8 & 3.7\\
LowResC & 397 & 2.1 & 4.6\\
LowMassC & 0 & - & 7.8\\
\hline
HighSoftMW & 39 & - & 1.5\\
MedSoftMW & 1911 & 4.0 & 1.1\\
LowSoftMW & 3942 & 4.5 & 1.1\\
~ & 1766 & 22.4 & 0.3\\
LowFloorMW & 4514 & 8.8 & 1.0\\
LowResMW & 1576 & 2.5 & 2.0\\
LowViscMW & 3639 & 4.0 & 1.0\\
MedGasMW & 3765 & 3.6 & 0.5\\
HighGasMW & 4448 & 0.6 & 0.4\\
FlatMW & 4124 & 5.7 & 0.8\\
~ & 2237 & 11.3 & 0.3\\
HighResFlatMW & 4445 & 16.0 & 0.3\\
\hline
\end{tabular}
\caption{\label{visctimes}Mean viscous time-scales and simulation lengths for all runs for the time from the first to the last recorded interaction. These time-scales are the mean time-scales during the time period from the first to the last recorded interaction. time-scales are not given for LowMassC and HighSoftMW. There were no interactions in LowMassC, as it did not form clouds. Interactions were detected in HighSoftMW, but only in clumps within the central bar, which do not contribute to disc viscosity. The viscous time-scale for the first $300$ Myr of LowSoftMW and FlatMW are also given for more direct comparison with HighResFlatMW. } 
\end{table}

\begin{figure*}
\includegraphics[width=1.\columnwidth]{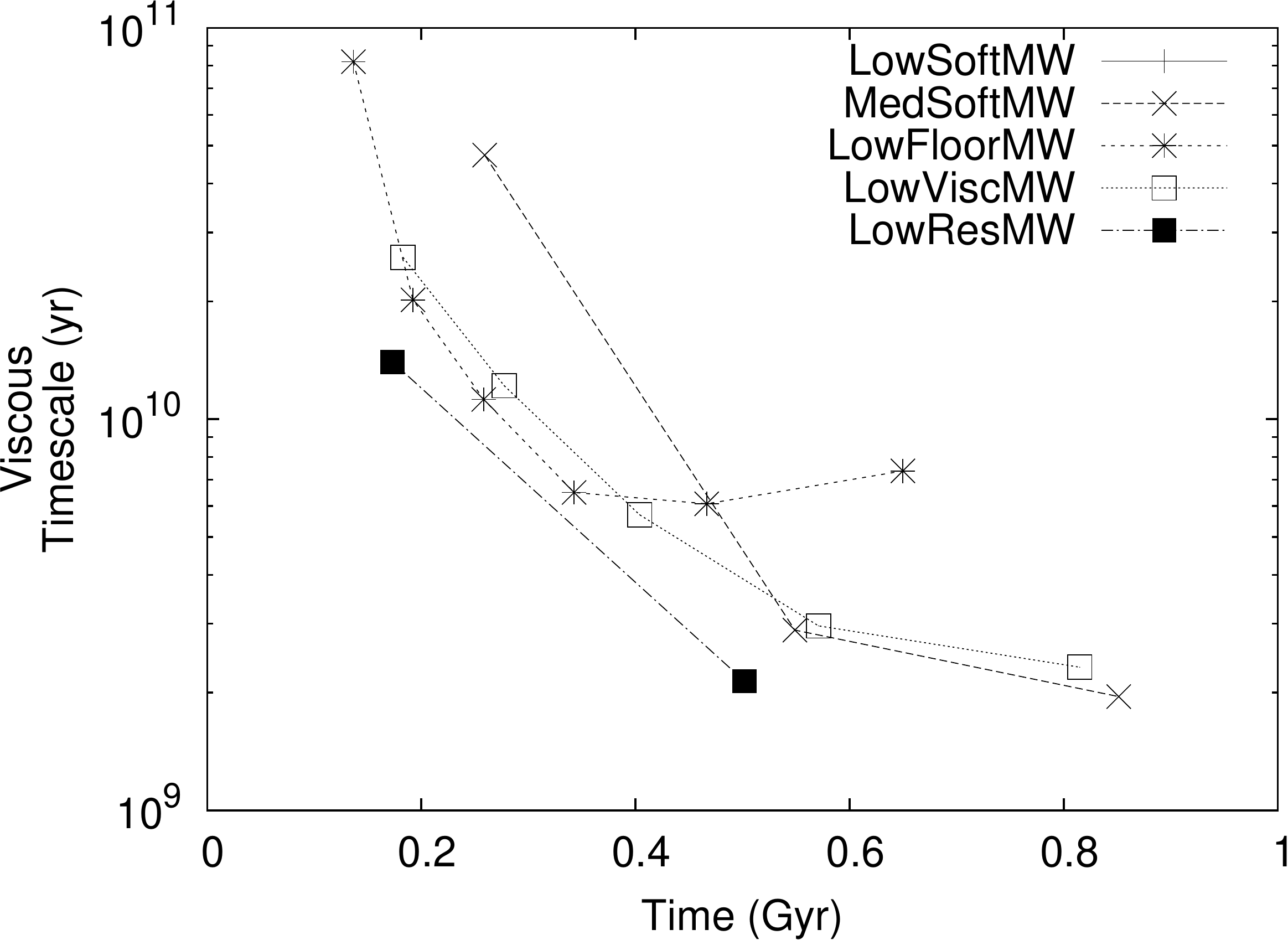}~\includegraphics[width=1.\columnwidth]{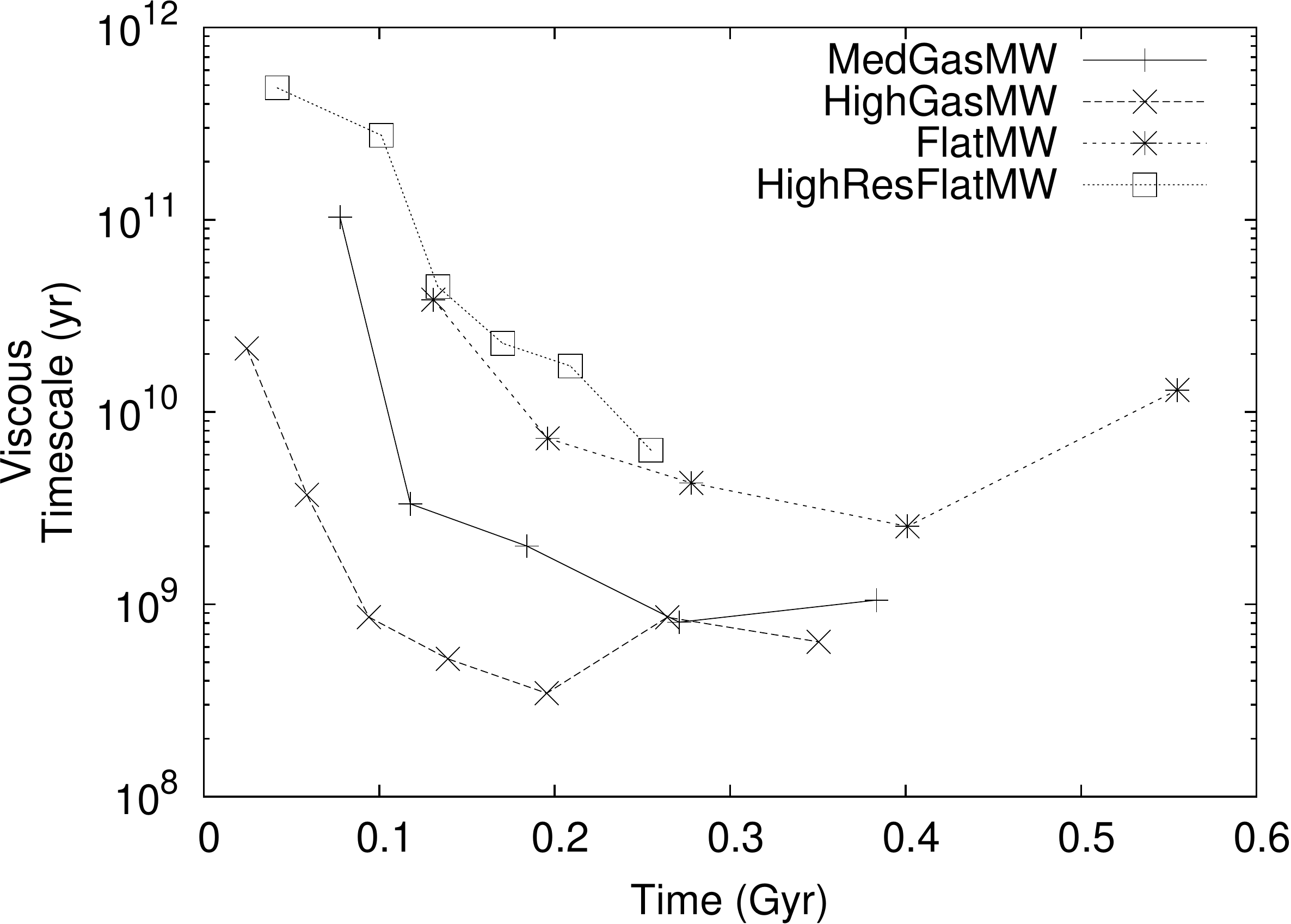}\\
\caption{\label{visctimedisc} Viscous time-scales for disc models that ran for $>800$ Myr (left) and $\le 800$ Myr (right). At early times, some models give negative time-scales, but as these values are large, they are not as dynamically important and are not plotted.}
\end{figure*}

The viscous time-scale is calculated using the method described in section \ref{energysection} and is plotted in Fig.~\ref{visctimedisc}. Each point is calculated from $600$ collisions. There is a general trend 
toward lower time-scales as the simulation evolves, and the final time-scales are generally below $10$ Gyr, with many approaching $1$ Gyr. This decreasing trend coincides with a trend of the number of clouds lowering 
and the mass of 
individual 
clouds increasing. The time-scales are less than a Hubble 
Time, and so should have some 
significant effect on the evolution of a galaxy, contrary to the predictions of \bellcitedot

This energy loss is seen in the mean specific kinetic energy of the gas in LowSoftMW, dropping from $1.9\times10^{14}$ erg/g at $t=170$ Myr to $1.1\times10^{14}$ erg/g at $t=1010$ Myr.  As expected, this loss is primarily in the clouds - the diffuse gas only drops from $1.9\times10^{14}$ erg/g to $1.7\times10^{14}$ erg/g in the same time period. We can make an additional crude estimate of the total viscous time-scale, $t_{\nu}=(t_2-t_1) k_1/(k_1-k_2)$, where $t_i$ and $k_i$ are the time and specific kinetic energy at these two outputs. This gives a time-scale of $t_\nu\sim2$ Gyr, around half of the value of $t_\nu=4.5$ Gyr from our method in section \ref{energysection}. This is either because additional energy is being dissipated through internal processes in clouds, or because interactions are being missed by our interaction finding procedure. The energy loss of the gas over this time period in LowViscMW is the same to a precision of less than $0.1\times10^{14}$ erg/g, suggesting that this additional energy loss is `physical' and not dominated by artificial viscosity. Nevertheless, we can not evaluate how much of this additional energy loss is directly due to cloud-cloud collisions, and so it is more informative to use the procedure in \ref{energysection} to calculate the viscous time-scale.

The mean viscous time-scales 
from all interactions over each entire simulation for both the Milky Way 
and collapse models are tabulated in Table~\ref{visctimes}. Despite the variation of parameters, many  of the time-scales are within a narrow range, from $3$--$5$ Gyr. Modifying the artificial viscosity (LowViscMW) did not appear to significantly change the viscous time-scale. The softening length in HighSoftMW ($600$ pc) was large enough to completely quench cloud formation, except for a few clumps that formed within the central 
bar instability. We do not include a viscous time-scale here as the 
mechanisms for formation and interaction are different to those of 
molecular clouds in nearly circular orbits. Feedback processes from star 
formation and AGN would also be more important here than in the other 
models. However, lowering the softening length from $100$ pc to $60$ pc (MedSoftMW to LowSoftMW), while increasing the number of clouds produced, did not significantly alter the viscous time-scale.

\begin{figure}
\includegraphics[width=1.\columnwidth]{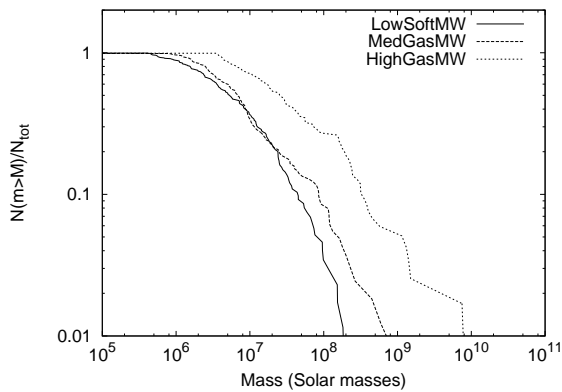}\\
\caption{\label{spectgas} Cumulative cloud mass spectra across runs with varying gas fraction.}
\end{figure}

HighGasMW has a significantly shorter viscous time-scale at $0.6$ Gyr, and indeed there appears to be a trend of decreasing viscous time-scale with increasing gas fraction. This is clearer if we compare the models over the same time period. The viscous time-scale over the first $430$ Myr is $7.1$ Gyr for LowSoftMW, $1.5$ Gyr for MedGasMW, and $0.6$ Gyr for HighGasMW. Increasing the gas fraction increases the mass of the cloud population (Fig.~\ref{spectgas}), which increases the frequency and dissipative efficiency of collisions.

HighResFlatMW is our highest resolution simulation, but has different initial conditions to LowSoftMW due to the more stringent stability requirements at high resolution (detailed in section \ref{mwics}). The flat discs of FlatMW and HighResFlatMW caused cloud formation to occur earlier than in LowSoftMW. A resolution dependence is also evident: The $2.5$x increase in mass resolution from FlatMW to HighResFlatMW caused a $1.4$x increase in the viscous time-scale, and the $5$x increase in mass resolution from LowResMW to LowSoftMW caused a $1.8$x increase in the viscous time-scale.

\begin{figure*}
\includegraphics[width=1.\columnwidth]{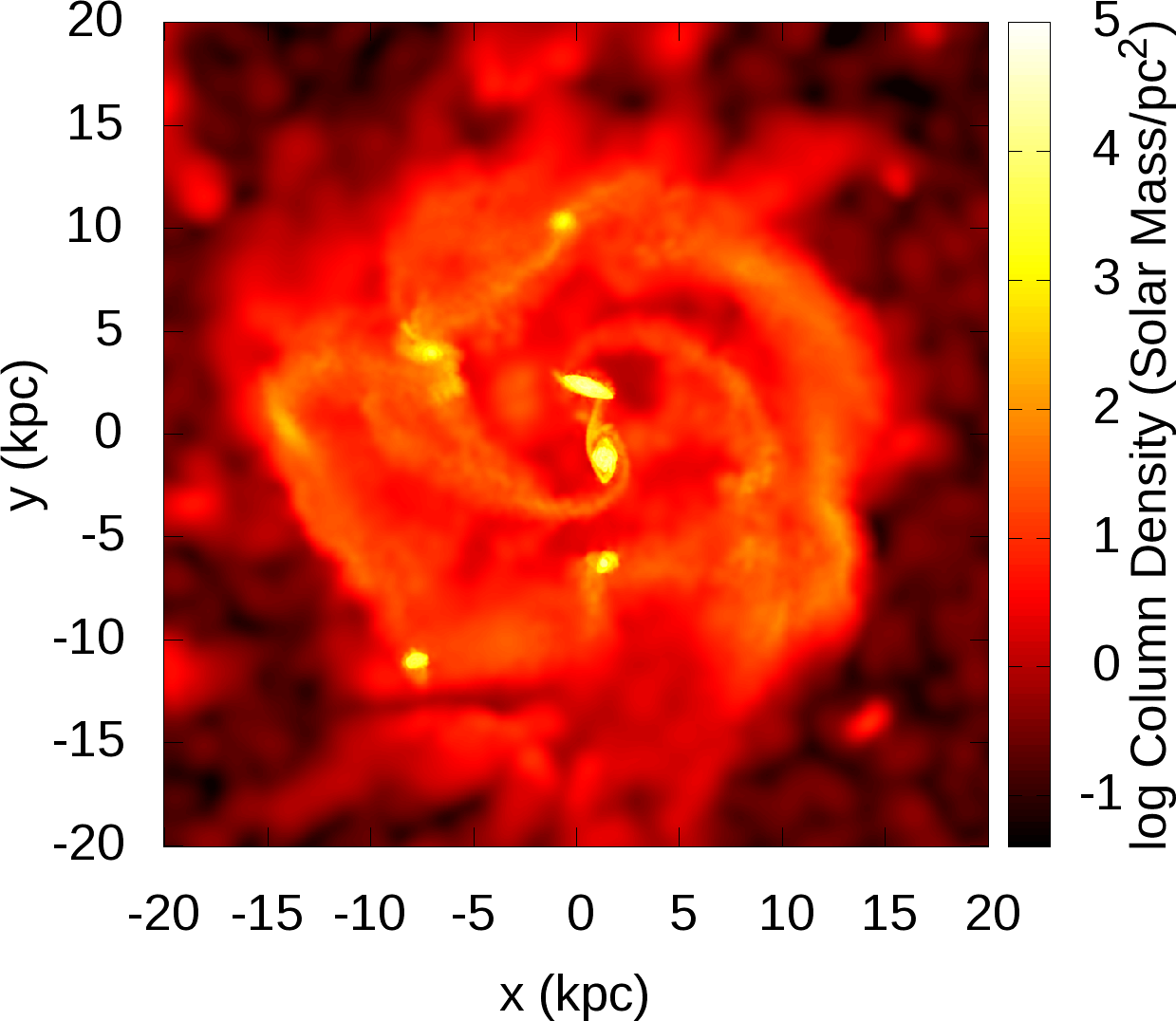}~\includegraphics[width=1.\columnwidth]{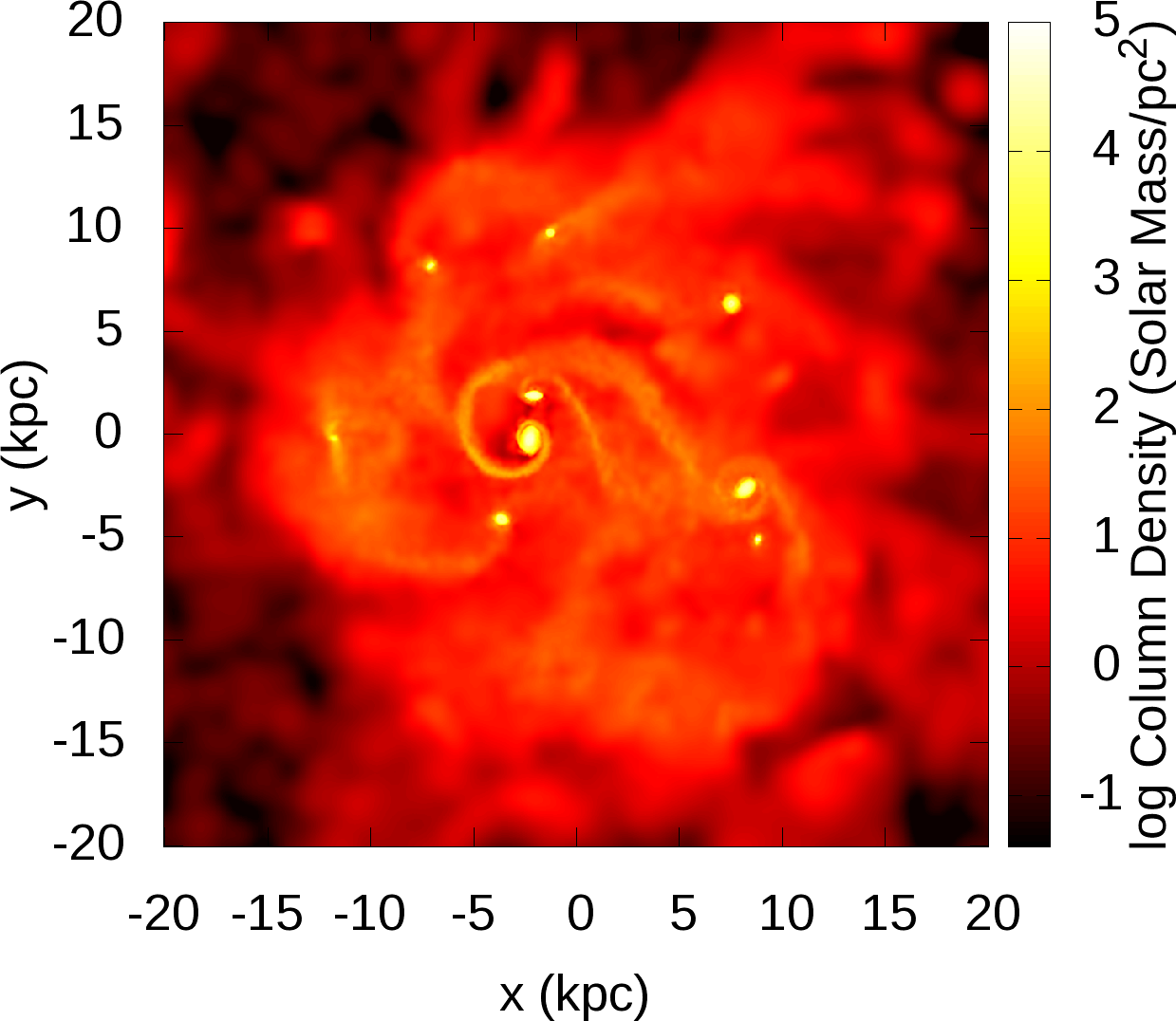}\\~\\
\includegraphics[width=1.\columnwidth]{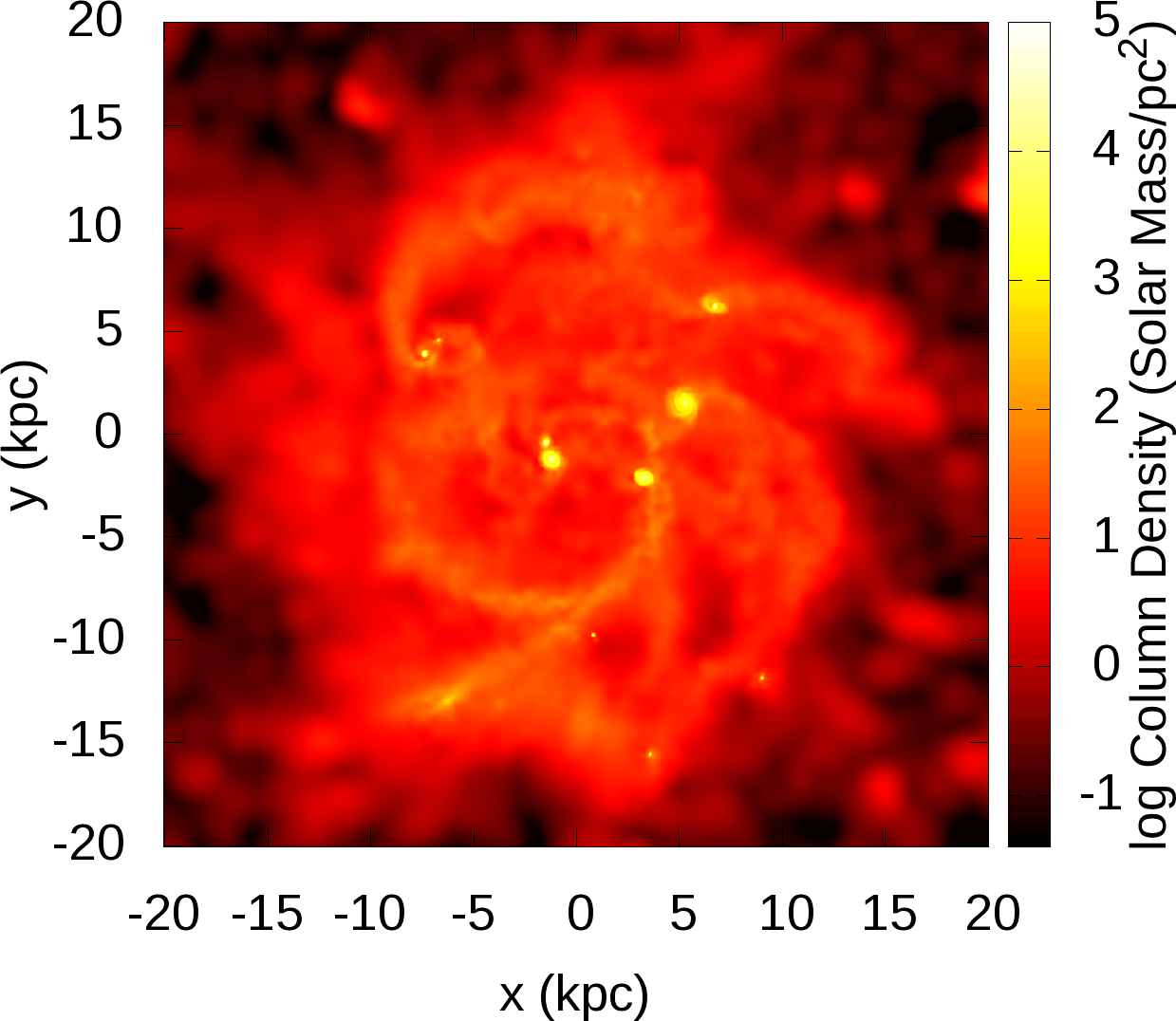}~\includegraphics[width=1.\columnwidth]{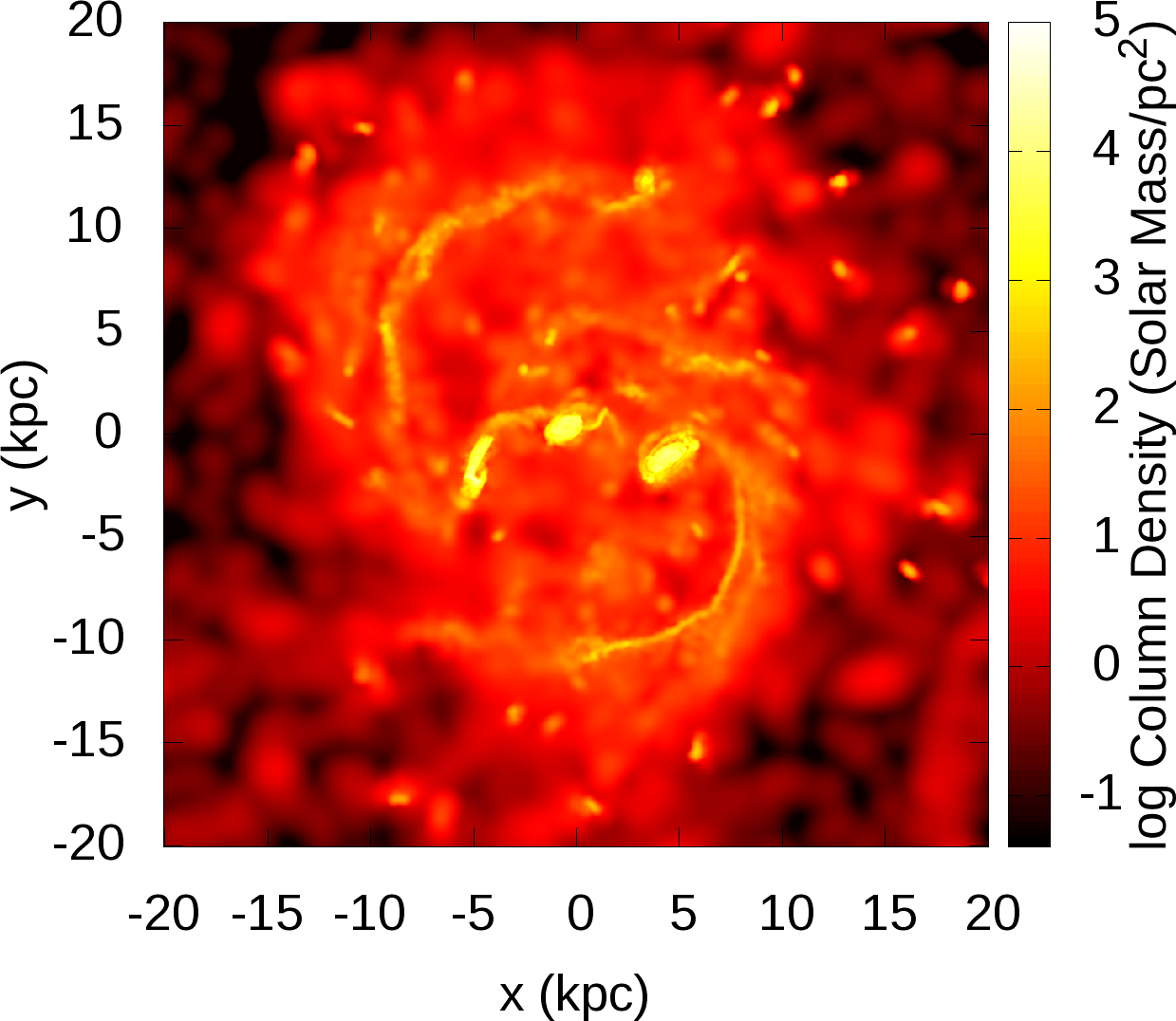}\\
\caption{\label{softdiscs} Impact of varying the softening length and
resolution in collapse runs at t=$3.5$ Gyr. Top left is HighSoftC ($514$ pc,
$3\times10^4$ K), top right is MidSoftC ($200$ pc, $3\times10^4$ K),
bottom left is LowSoftC ($60$ pc, $3\times10^4$ K) and bottom right is
LowResC ($60$ pc, $300$ K). Although HighSoftC, MidSoftC and LowSoftC produce different numbers of clouds initially (more clouds for a shorter softening length), after $\sim 500$ Myr of collisions all three models have $\sim7$ large clumps. Despite the low temperature floor, the limited resolution of LowResC produces an unstable disc, instead of a swam of dense clumps as in LowSoftFloorC.}
\end{figure*}

\subsection{Monolithic Collapse Model}

\begin{figure}
\begin{center}
\includegraphics[width=1.\columnwidth]{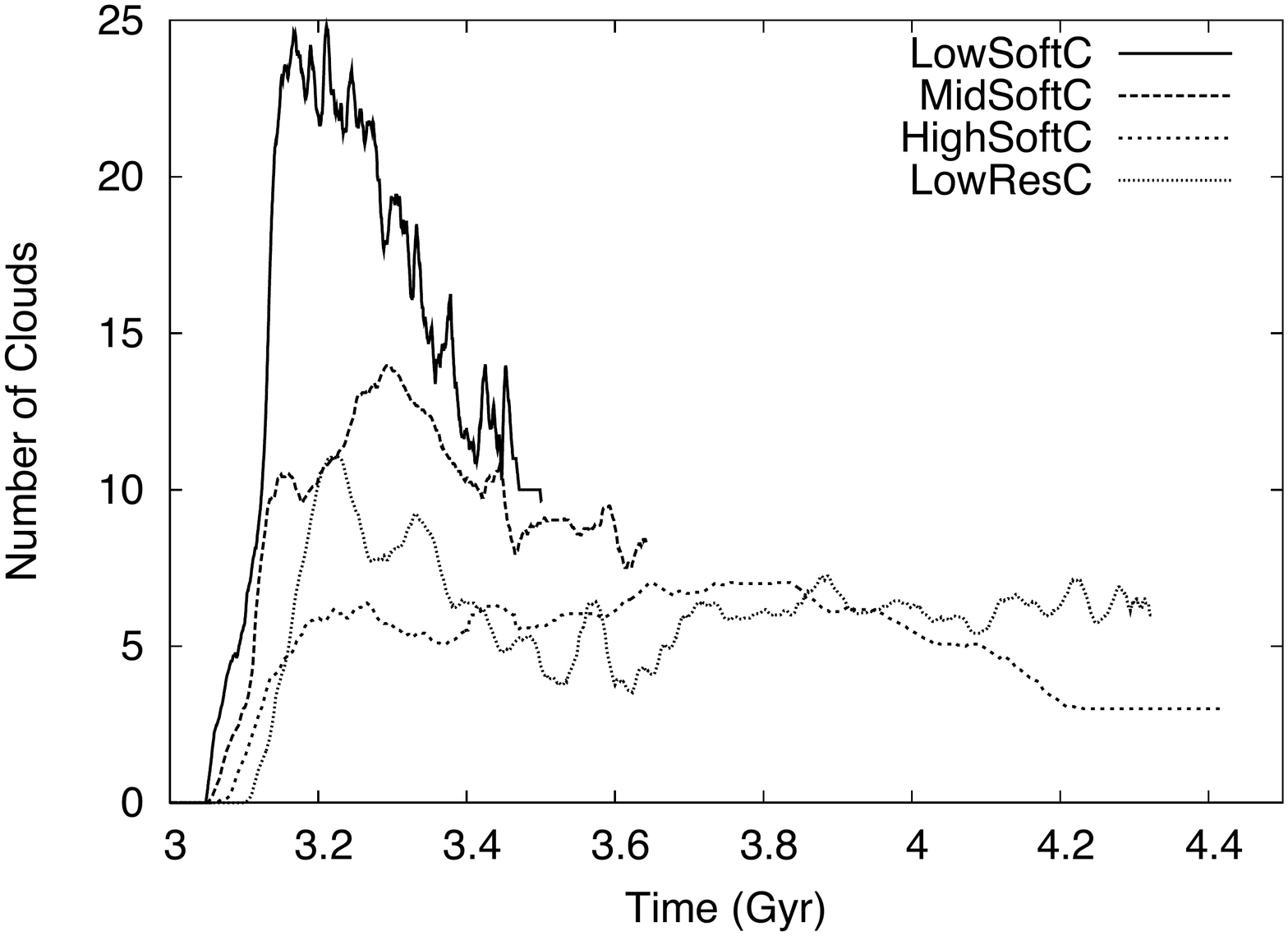}
\end{center}
\caption{\label{nccollapse} Number of clouds in collapse models 
(excepting LowSoftFloorC). To smooth the data, each plotted point is an average of the $29$ data points centred on it. Being very unstable, these systems formed a 
few large clumps rather than many small clumps.}
\end{figure}

In all models the gas collapse proceeds as soon as cooling is turned on, 
thus breaking the hydrostatic equilibrium. The High-S profile slowed the 
collapse sufficiently for the infalling gas to fragment into clouds at a 
large radius, although these clouds are are too diffuse to be found by 
the cloud identification algorithm. As the simulation progresses, these 
clouds start to merge (from $t\sim3$ Gyr in all runs except for LowMassC), and 
reach the effective threshold density of our cloud-finder. The 
number of clouds quickly reaches a maximum (see Fig.~\ref{nccollapse}).
These clouds combine to form a disc. The number and size of clouds these discs fragment into varies greatly between our models.

\begin{figure*}
\includegraphics[width=1.\columnwidth]{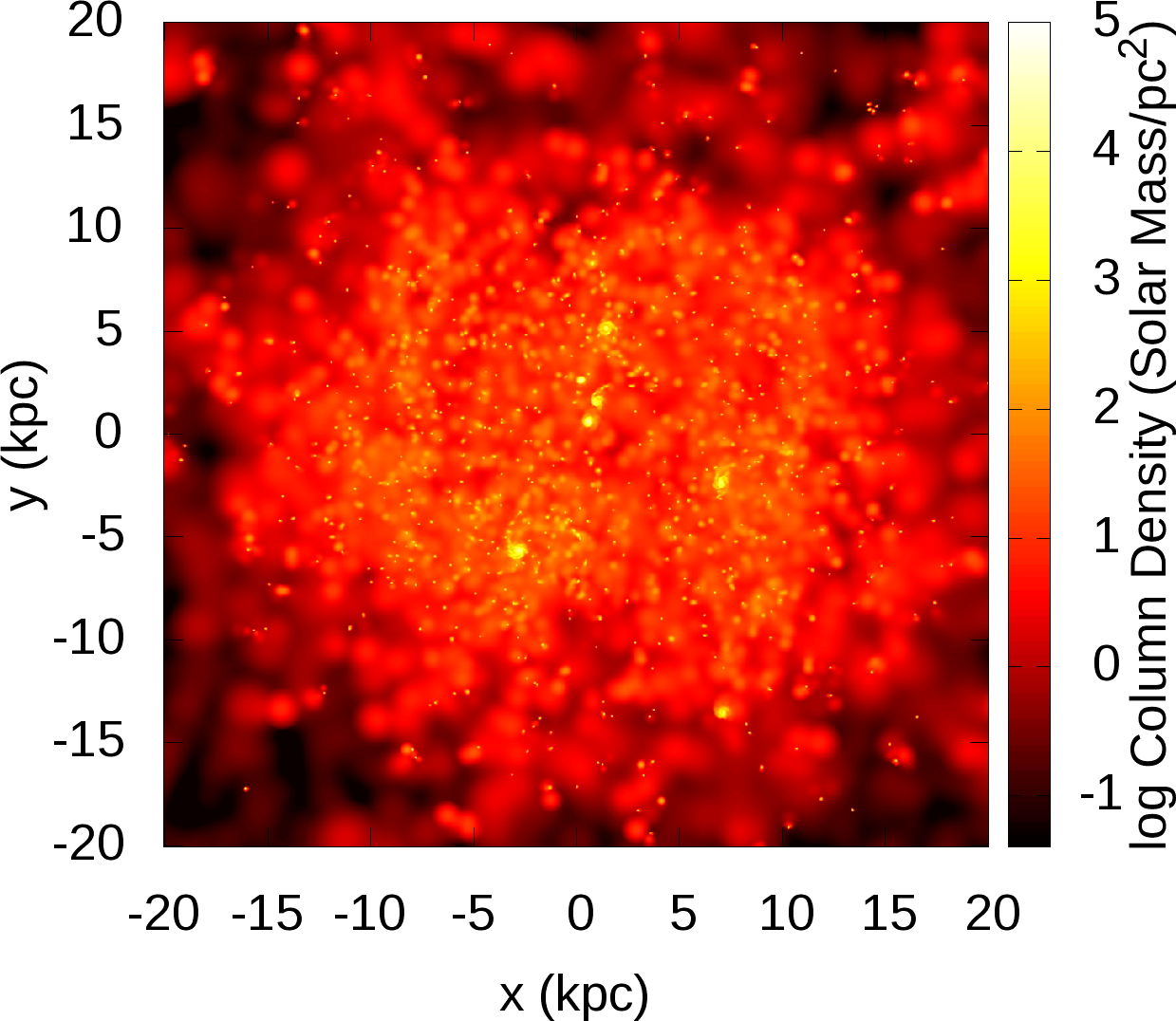}~\includegraphics[width=1.\columnwidth]{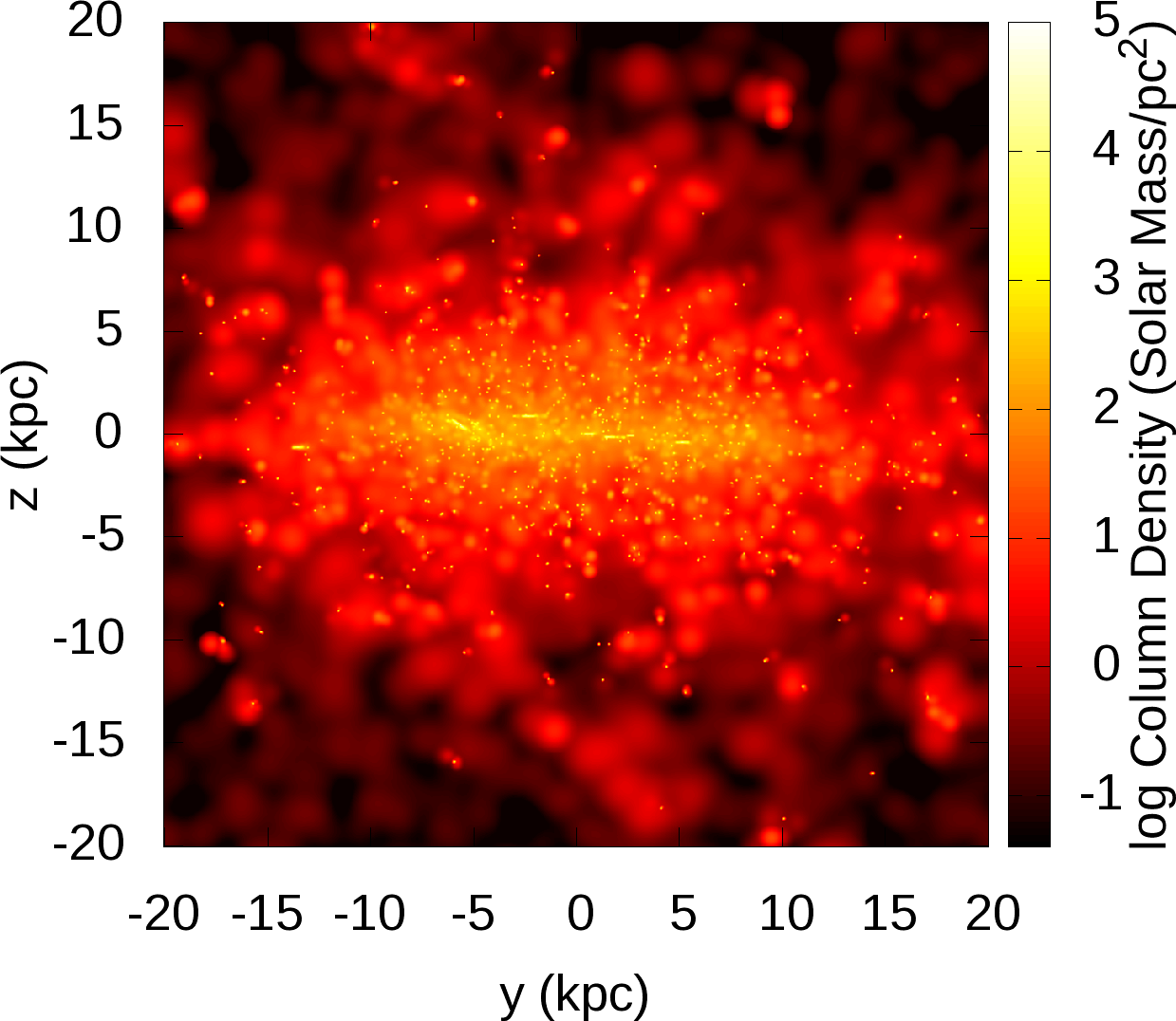}\\
\caption{\label{clumpswarm} Face-on and side-on density plots of
LowSoftFloorC at t=$3.7$ Gyr. The swarm of clumps has a half-mass height of $7.8$ kpc. The disc is very chaotic: $10$ kpc, the azimuthally averaged tangential velocities and velocity dispersions are $180$ km s$^{-1}$ and $105$ km s$^{-1}$.}
\end{figure*}

In HighSoftC, MidSoftC, LowSoftC and LowResC, the disc is extremely 
unstable, collapsing into $\sim7$ massive (several times $10^9 M_\odot$ 
in mass) clumps (Fig.~\ref{softdiscs}). These are not 
small-scale GMC-style clumps as found in the Milky Way simulations, and 
perhaps this level of collapse is more 
analogous to the gas-rich clump-cluster galaxies found at high redshift 
\citep[\eg][]{2005ApJ...627..632E}. In the simulations of 
\citet{2007ApJ...670..237B} and \citet{2009ApJ...703..785D}, the large 
clumps in clump-cluster galaxies coalesce into a central bulge, forming 
a more stable disc. These simulations differ to ours particularly in 
that they include star-formation and feedback. With infalling material, 
\citet{2009ApJ...703..785D} finds the clumpy phase can last for several 
Gyr. 

The heavy clustering in this discs dictated that they could only be 
evolved for $<1$ Gyr after formation 
(which takes $\sim3$ Gyr) due to problems with the SPH solver.
The high densities cause a large 
increase in the number of particles with smoothing lengths at the 
minimum allowed which contributes to an ${\cal O}(n^2)$ slowdown.
 
The simulations of \citet{2009MNRAS.396..191K}, while including star-formation (but not explicit feedback), also produce a disc with large-scale gravitational instabilites. Both our and Kaufmann's 
models have a temperature floor of $3\times10^4$ K, as a very crude form of feedback. Including star-formation and more self-consistent feedback method could produce a stable disc \citep{2006MNRAS.373.1074S, 2010ApJ...717..121C}.

In LowSoftFloorC, the low temperature floor allows the halo clouds to 
condense into dense ($n\sim 10^{4}-10^{5}$cm$^{-3}$) clumps 
(Fig.~\ref{clumpswarm}). Their low 
cross-section means that their coalescence has properties of a  
collisionless collapse. So in addition to an unstable disc, there exists 
a swarm of clumps with a half-mass height of $7.8$ kpc. Their 
ellipsoidal distribution and high 
densities are reminiscent of globular clusters, but the inclusion of 
feedback would definitely increase the cloud cross-sections and produce
a more dissipated and flattened disc.

\begin{figure}
\begin{center}
\includegraphics[width=1.\columnwidth]{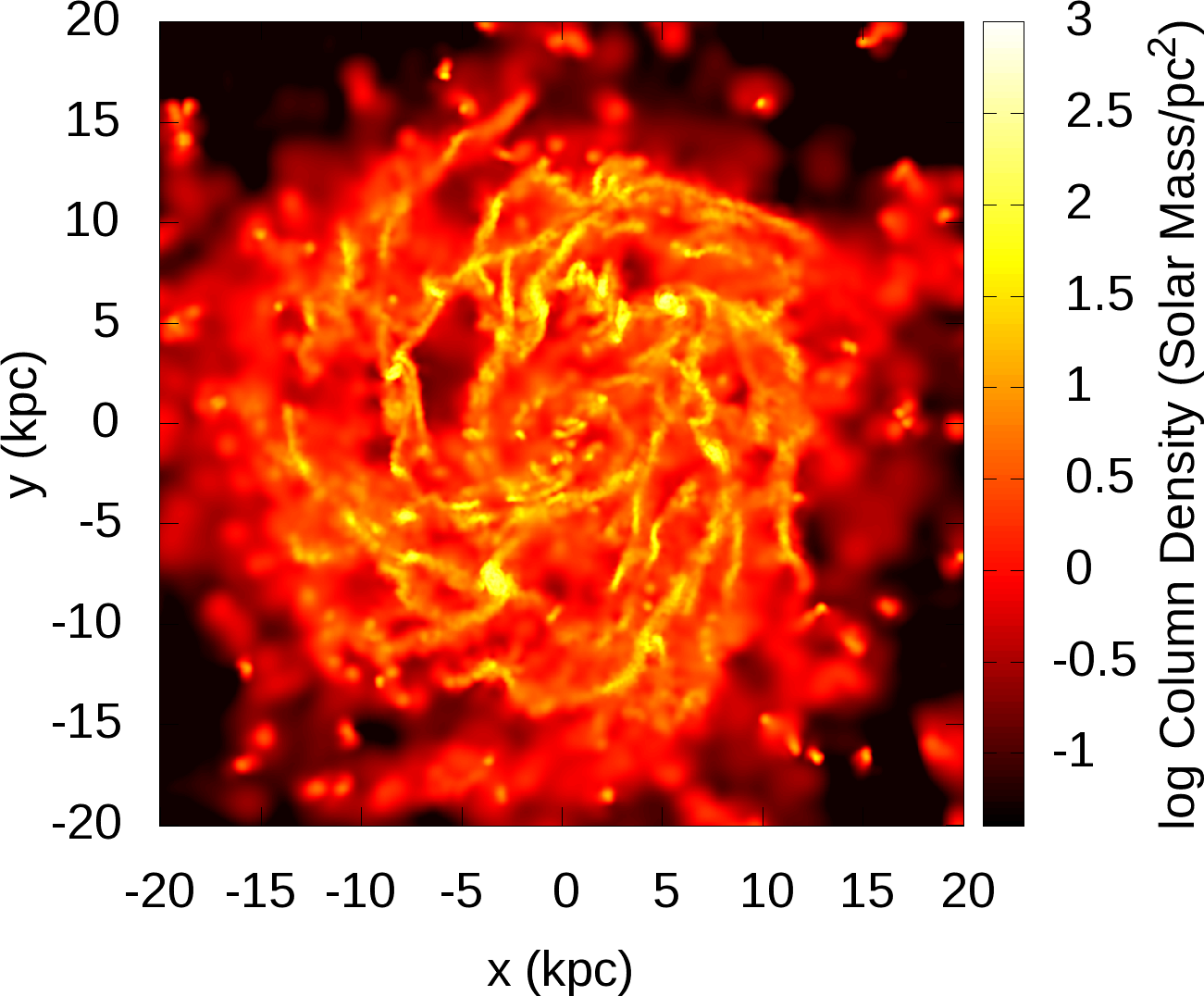}\\
\end{center}
\caption{\label{stablediscC} LowMassC at $t=6.0$ Gyr. The disc undergoes spiral instabilities but does not fragment into clumps as the other collapse models do.}
\end{figure}

LowMassC is the only run that produces a disc that does not collapse 
into large clumps (Fig.~\ref{stablediscC}), although it took 
considerably longer to form ($\sim4.5$ Gyr) and the disc is still 
dominated by spiral instabilities. Discs are unstable to bar formation when the disc mass fraction is greater than the spin parameter ($m_d>\lambda_{\rm G}$) \citep{1982MNRAS.199.1069E,2008MNRAS.386.1821F}, so a lower mass disc is more stable. 
If the bar is too strong, it may fragment into large clumps. This 
instability may well drive the infalling clouds into a few large clumps 
in the 
higher mass models.

As seen in Table~\ref{visctimes}, the viscous time-scales for the 
collapse runs trend toward lower values than the Milky Way simulations - 
around $1-2$ Gyr. Though the number of 
interactions is not as large as in the Milky Way models, they occur over 
a short period (e.g. all $566$ interactions in LowSoftC are within 
$\sim500$ Myr). The number of clouds is small, so each cloud undergoes 
many collisions, producing a short viscous time-scale.

\subsection{Comparison with Analytical Model}

\bellcite argued that while cloud collisions are not uncommon (occurring $\apgt1$ time per orbit), the 
low efficiency of cloud collisions produces a long viscous time-scale. This efficiency is measured with a parameter 
$\eta$, equal to the fraction of a cloud's energy that is lost in a 
collision (not entirely dissimilar from a coefficient of restitution). When two clouds merge completely, the fraction of kinetic energy lost is well approximated by
$\eta=(v_{\rm rel}/v_{\rm rot})^2$, 
where $v_{\rm rel}$ is the relative velocity of the clouds, and $v_{\rm rot}$ is their rotational velocity which is roughly constant for a galaxy. This is consistent with our numerical results. The analytical model of \bellcitecom finds that $\eta\lesssim10^{-2}$ for a Milky-Way-like model, concluding that cloud-cloud collisions are not an efficient sink of energy, with $t_\nu\sim1000$--$2000$ Gyr. 

The complex interactions that occur between clouds in our simulation mean that it is not straightforward to 
determine values for $\eta$. Several of our merger and separation events can take place within what is
really a single extended interaction, which lowers the average time between interactions significantly. Indeed, we find 
the interaction rates are on the order of one separation or merger 
event per cloud every $50-60$ Myr for LowSoftMW, MedSoftMW, LowFloorMW, LowViscMW, HighSoftC and LowMassC. The greatest interaction time-scale was in LowSoftFloorC ($335$ Myr), and the smallest was in LowSoftC ($14$ Myr).

It is difficult to track the number of interactions over a full merger process, as additional clouds often interact with the merging clouds. We carefully examined examined a span of time around each of a sample of $10$ recorded interactions in LowSoftMW on an iteration-by-iteration basis to determine the number of recorded interactions per `real' interaction. These interactions were selected so that they were evenly distributed across the simulation ($\sim2$ every $5000$ iterations). We initially examined a period of $\pm800$ iterations around the interaction, and if no `real' interaction was observed during this time, this was extended to $\pm2000$ iterations. Several different behaviours were observed:

\begin{itemize}
\item In two cases, no real interactions were observed; outer parts of a cloud were attaching and detaching to the main cloud, and dissolving and condensing across the cloud density threshold, causing a number of recorded interactions which did not correspond to any clear long-term merger, scattering or separation event.
\item Three events were `messy' interactions with $6$, $7$ and $16$ recorded interactions per real event; the event consisting of $16$ recorded split and merge events was a scattering event where the clouds passing by each other several times before separating for a final time.
\item Four more events were more `tidy' interactions, with 1, 2, 3, and 4 interactions per real event.
\item The last event was a series of mergers in rapid succession - $3$ recorded and $3$ `real' mergers.
\end{itemize}

Overall, there was a mean of $4.9$ recorded interactions per examined period, with a standard deviation of $4.3$. A total of $11$ `real' interactions were observed, giving $4.5$ recorded interactions per real interaction. This increases our interaction time-scale to one event per $\sim250$ Myr for the LowSoftMW-like models. This is approximately once per orbit at a solar radius. The analytic estimate in \bellcite of the cloud-cloud collision rate is $\sim100$ Myr, which is of similar order.

\begin{figure*}
\includegraphics[width=2.\columnwidth]{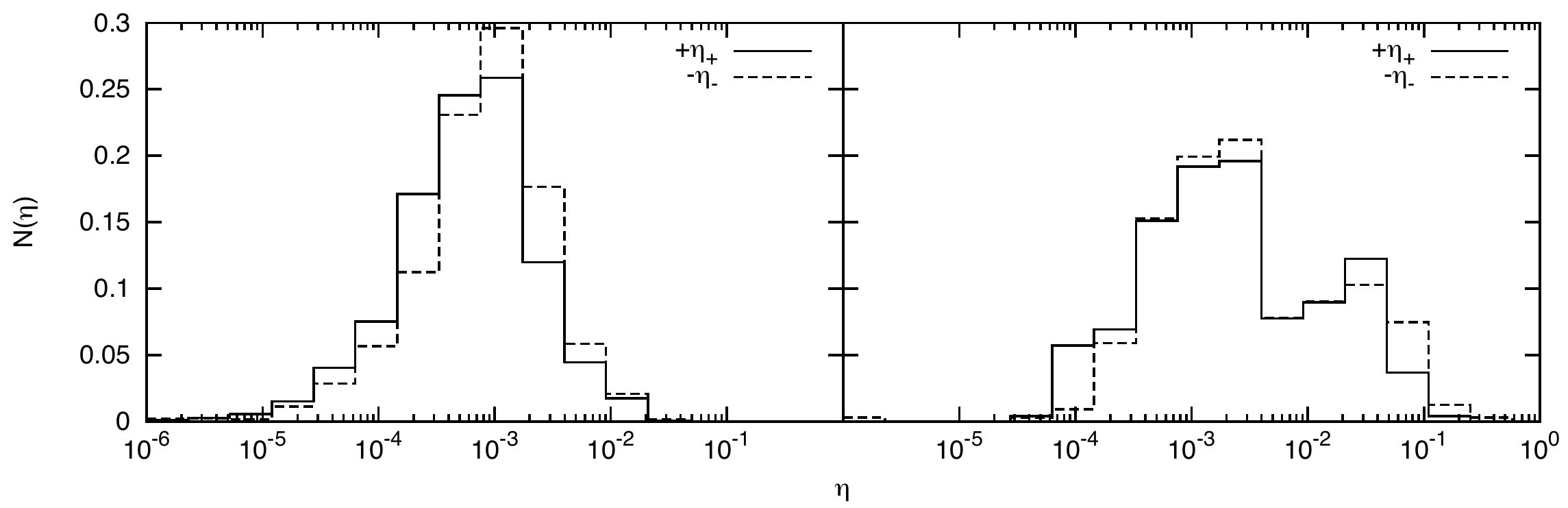}\\
\caption{\label{etaplots} Distributions of the fraction of energy lost in a collision $\eta$, in bins of 0.35 dex. Left: LowSoftMW, Right: LowSoftC. For each simulation, the distribution of all $\eta_-<0$ and $\eta_+>0$ are plotted separately. In both cases, the median value of $|\eta_-|$ is greater than the median value of $|\eta_+|$, even though both models show a positive viscous time-scale.}
\end{figure*}

We can estimate an $\eta$ for the interactions in our models by 
\begin{equation}
\eta=-(\Delta K+\Delta \phi)/(K_c),
\end{equation}
 where $\Delta K$ and $\Delta \phi$ are the change in kinetic and potential energy of a cloud, and 
$K_c$ is the total kinetic energy of both clouds before collision. $\eta$ can be negative, as energy is converted from internal motions into orbital kinetic energy during separations. The clouds all have similar velocity because of the flat rotation curve, so the total energy lost is primarily dependent on $\eta$ and the cloud masses. We find for most interactions $|\eta|$ is on the order of 
$\sim0.002$ (Fig.~\ref{etaplots}). If we separate our $\eta$ values into two sets, $\eta_-$ for $\eta<0$ and $\eta_+$ for $\eta>0$, we find that the median value of $|\eta_-|$ is greater than the median value of $|\eta_+|$, even though the viscous time-scale is positive. This is because although $\eta$, the {\em relative} energy change is larger for interactions which increase orbital energy ($\eta_-<0$) than those which decrease orbital energy ($\eta_+>0$), the {\em absolute} change in energy is larger for interactions which decrease orbital energy than {\em increase} it - \ie these interactions tend to occur between clouds with greater mass. Although it is not apparent on these plots, there are several collisions for which $\eta$ is very large, with $\eta>0.1$. These interactions occurred within the $1$ kpc of galaxy centre, and only after $\sim400$ Myr. These are clouds that have been strongly scattered by interactions and fallen down the potential well, colliding with speeds of $>100$ km s$^{-1}$.

Our interactions are no more efficient at removing energy than in \bellcitecom and are no more common, yet the \bellcite model predicts $t_\nu\sim1000$--$2000$ Gyr, while our simulations have $t_\nu<10$ Gyr. Our simulated discs are more energetic than standard Milky Way models: the velocity dispersion in LowSoftMW is $\sim20$ km s$^{-1}$ at $7.5$ kpc, more than triple the standard Milky Way value used in \bellcite ($6$ km s$^{-1}$). However, this is not the cause of the large difference between the model of \bellcite and our own. Here we derive our own model for $\eta$, and contrast this with the model in \bellcite to find the source of this disparity.


We can split the velocity components of $v_{\rm rel}$ into tangential and radial components to give

\begin{equation}
\eta\sim\frac{v_{\rm rel}}{v_{\rm rot}}=\frac{R^2 (\dot{\phi}_1-\dot{\phi}_2)^2 + (\dot{R}_1-\dot{R}_2)^2}{v_{\rm rot}^2}.
\end{equation}

If we make the epicyclic approximation\citep{2008gady.book.....B}, that the deviation from a circular orbit is small compared to the radius of the orbit ($R=R_{\rm g}+x$, where $R_{\rm g}$ is the `guiding centre' of the orbit, and $x\ll R$ is the radial excursion), then $\dot{R}=\dot{x}=X \kappa \cos(\kappa t + \alpha)$, (where $X$ is the maximum radial excursion of a cloud, $kappa$ the epicyclic frequency, and $\alpha$ is a phase parameter) and $\dot{\phi}=R_{\rm g} v_{\rm rot}/R^2$ from conservation of momentum in a flat rotation curve. Hence $R^2(\dot{\phi}_1-\dot{\phi}_2)^2=(v_{\rm rot}^2/R^2)(R_{\rm g,1}-R_{\rm g,2})^2$ - the tangential component of the difference in velocity depends only on the radial distance between the clouds' guiding radii.

The radial component is more difficult to calculate, as it depends on the phase of the interaction. We can estimate the maximum $\eta$ by assuming the clouds are perfectly out of phase - that is, 

\begin{eqnarray}
(\dot{R}_1-\dot{R}_2)^2 \sim (X_1 \kappa_1 + X_2 \kappa_2)^2 &\sim& 2 v_{\rm rot}^2 \left(\frac{X_1}{R_{\rm g,1}} + \frac{X_2}{R_{\rm g,2}}\right)^2 \nonumber\\
&\sim& \frac{2v_{\rm rot}^2}{R^2} (X_1 + X_2),
\end{eqnarray}
as $R\sim R_{\rm g}$. For clouds to collide precisely out of phase, they must have the same guiding radius, and so $R_{\rm g,1}-R_{\rm g,2}=0$. Hence, if $X_1\sim X_2\sim X$, then $\eta_{\rm max,r} = 8 X^2/R^2$. If the clouds are at their maximum deviation when they collide, then their radial velocities are zero, but their relative $\phi$ velocities are maximised, that is, $\dot{\phi}_1-\dot{\phi}_2=2X$, and so $\eta_{{\rm max},\phi} = 4 X^2/R^2$. These coefficients give the maximum $\eta$, but we should nevertheless expect $\eta\sim X^2/R^2$, i.e. $\eta$ depends on the radial excursion of clouds.

This can also be expressed in terms of a velocity dispersion. We can calculate the velocity dispersion by

\begin{equation}
v_s^2 = \langle(\mathbf{v}-\mathbf{v_{ave}})^2\rangle = \langle(\dot{x})^2\rangle + \langle R^2(\dot{\phi}-\Omega_g)^2\rangle
\end{equation}

Assuming a flat rotation curve and that $X$ and $\kappa$ are more or less constant within the region of interest, the radial component is $\dot{x}=X \kappa \cos(\kappa t + \alpha)$ , hence
 
\begin{equation}
\langle(\dot{x})^2\rangle=(1/2)X^2 \kappa^2=X^2 v_{\rm rot}^2/R^2,
\end{equation}
and the tangential component is $R(\dot{\phi}-\Omega_g)=-2X \Omega_g \sin(\kappa t + \alpha)$, hence 

\begin{equation}
\langle R^2(\dot{\phi}-\Omega_g)^2\rangle=2X^2 \Omega_g=2 X^2 v_{\rm rot}^2/R^2.
\end{equation}

This gives 

\begin{equation}\label{vsX}
v_s^2 = 3 v_{\rm rot}^2 (X^2/R^2),
\end{equation} and so

\begin{equation}
\eta \sim \frac{v_s^2}{v_{\rm rot}^2}.
\end{equation}

From these expressions for $\eta$ we can determine the dissipative time-scale from $t_\nu = t_c/\eta$.


We next summarise the model of \bellcitedot  In the limit of rapid collisions, the kinematic viscosity due to cloud-cloud collisions can be modelled as a Reynolds stress and expressed as $\nu\sim\lambda_d v_s$ \citep{1995fdp..book.....F}, where $v_s$ is the velocity dispersion, and $\lambda_d$ is the mean free path. The mean free path is $\lambda_d=v_s t_c$, where $t_c$ is the typical time between collisions. Similarly to our result, \bellcite states $\eta\sim \Delta R^2/R^2$, where $\Delta R$ is the radial distance between collisions. For the case of very rapid collisions, $\Delta R\sim\lambda_d$, so $\eta\sim\lambda_d^2/R^2$. This gives a viscous time-scale that should be equal to the dissipative time-scale,

\begin{equation}\label{tnutdisp}
t_\nu\sim \frac{R^2}{\nu}\sim \frac{R^2}{\lambda_d v_s} \sim t_c \frac{R^2}{\lambda_d^2} \sim \frac{t_c}{\eta}.
\end{equation}

Hence if we follow the description given in \bellcitecom the results should be equivalent to ours. Continuing to follow \bellcitecom we can set $t_c=M_{\rm cloud}h/(\Sigma_{\rm g} v_s\pi r^2_{\rm cloud})$, so
\begin{equation}\label{correctvisceq}
t_\nu=\frac{R^2}{v_s\lambda_d}\sim \frac{R^2}{v_s} \frac{\pi r_{\rm cloud}^2 \Sigma_{\rm g}}{M_{\rm cloud}h}.
\end{equation}

We can evaluate this using the Milky Way parameters of \bellcitecom that $r=7.5$ kpc, $v_{\rm rot}=220$ km s$^{-1}$, $v_s=6$ km s$^{-1}$, $\Sigma_{\rm g}=50 M_\odot$ pc$^{-2}$, $M_{\rm cloud}=10^5 M_\odot$, $h=100$ pc, and $r_{\rm cloud}=10$ pc to result in $t_\nu=14$ Gyr. However, \bellcite states $t_\nu\sim2000$ Gyr. This disagrees by a factor of $1/\eta$. It appears that \bellcite includes an additional factor of $\eta\sim0.008$ in the denominator - \ie $t_{\nu,Bell}\sim R^2/(\eta\nu)$. This $\eta$ is not necessary, as it is already included in the radial excursion or velocity dispersion, and as is clear from equation \ref{tnutdisp}, the expression $t_\nu\sim R^2/\nu\eta$ is not equivalent to the dissipative time-scale.

In \bellcite's rare collision case, $\nu\sim v_s \Delta R (t_\kappa/t_c)$, where $t_\kappa=2\pi/\kappa$ is the epicyclic time-scale. For a flat rotation curve $\kappa=\sqrt{2}\Omega\sim v_0/R$. The excursion $\Delta R$ is on the order of the radial excursion of the epicyclic motion of the clouds. \bellcite state $\Delta R\sim v_s/\kappa \sim v_s R/v_{\rm rot}$, which is consistent with our result in equation~\ref{vsX}. Putting this together gives
\begin{equation}
t_\nu \sim \frac{t_c/(2\pi) v_{\rm rot}^2}{v_s^2},
\end{equation}
\ie $\eta\sim 2\pi v_s^2/(v_{\rm rot}^2)\sim 0.023$. \bellcite uses a low surface brightness galaxy in this case, with $\Sigma_{\rm g}=10 M_\odot$ pc$^{-2}$ and $v_{\rm rot}=100$ km s$^{-1}$, which results in $t_\nu\sim23$ Gyr. Again, the value in \bellcite is much larger, $t_\nu\sim1000$ Gyr, which again is higher than our calculated value by a factor of approximately $1/\eta$.

These model are intended to apply in the limits of very frequent or very infrequent collisions where $t_c \Omega \ll 1$ or $t_c \Omega \gg 1$. In our simulations, we found that clouds 
collide about once per orbit, \ie $\Omega t_c\sim1$. However, we can contrast these results with those of \citet{1978Icar...34..227G}, who solve the 
Boltzmann equation for a system of inelastically colliding particles in a disc, and find for arbitrary $\Omega t_c$ that 
the viscosity is of order

\begin{equation}
\nu\sim v_s \lambda_d \frac{1}{1+(\Omega t_c)^2},
\end{equation} after we make the substitution that $\lambda_d\sim v_s t_c$. For $\Omega t_c=1$, $\nu=1/2 (\lambda_d v_s)$. The frequent collision case of \bellcitecom $\nu\sim v_s \lambda_d$, is accurate to this within an order of magnitude if we exclude the erroneous factor of $1/\eta$.


Substituting our typical cloud and disc parameters at $7.5$ kpc ($h\sim 25$ pc, $\Sigma_{\rm g}\sim 100 M_\odot/$pc$^{-2}$, $v_s\sim20$ km s$^{-1}$, $r_{\rm cloud}\sim 
35$ pc, and $M_{\rm cloud}\sim10^7 M_\odot$) for LowSoftMW at $t=1$ Gyr into this model gives a viscous time-scale of $1.1$ Gyr. This values somewhat underestimates our numerical results for the Milky Way models in Table~\ref{visctimes}, for most of which $t_\nu\ge4.0$ Gyr. The unstable disc of LowSoftC, forming from a collapse without stars, has very different properties at $R=7.5$ kpc, with $h\sim 250$ pc, $\Sigma_{\rm g}\sim 5000 M_\odot/$pc$^{-2}$, $v_s\sim100$ km s$^{-1}$, $r_{\rm cloud}\sim100$ pc, and $M_{\rm cloud}\sim10^9 M_\odot$. This gives $t_\nu=0.35$ Gyr, which agrees with our simulation result ($0.8$ Gyr) within a factor of $\sim2$. The analytical expression for $t_\nu$ was evaluated from order-of-magnitude arguments and assumptions that may not be entirely valid in our simulations - particularly in models with very few clouds, such as LowSoftC. Numerical factors also vary our simulation results by a factor of $\sim4$. Given these issues, it is not surprising that the agreement is not exact.

\begin{figure}
\begin{center}
\includegraphics[width=1.\columnwidth]{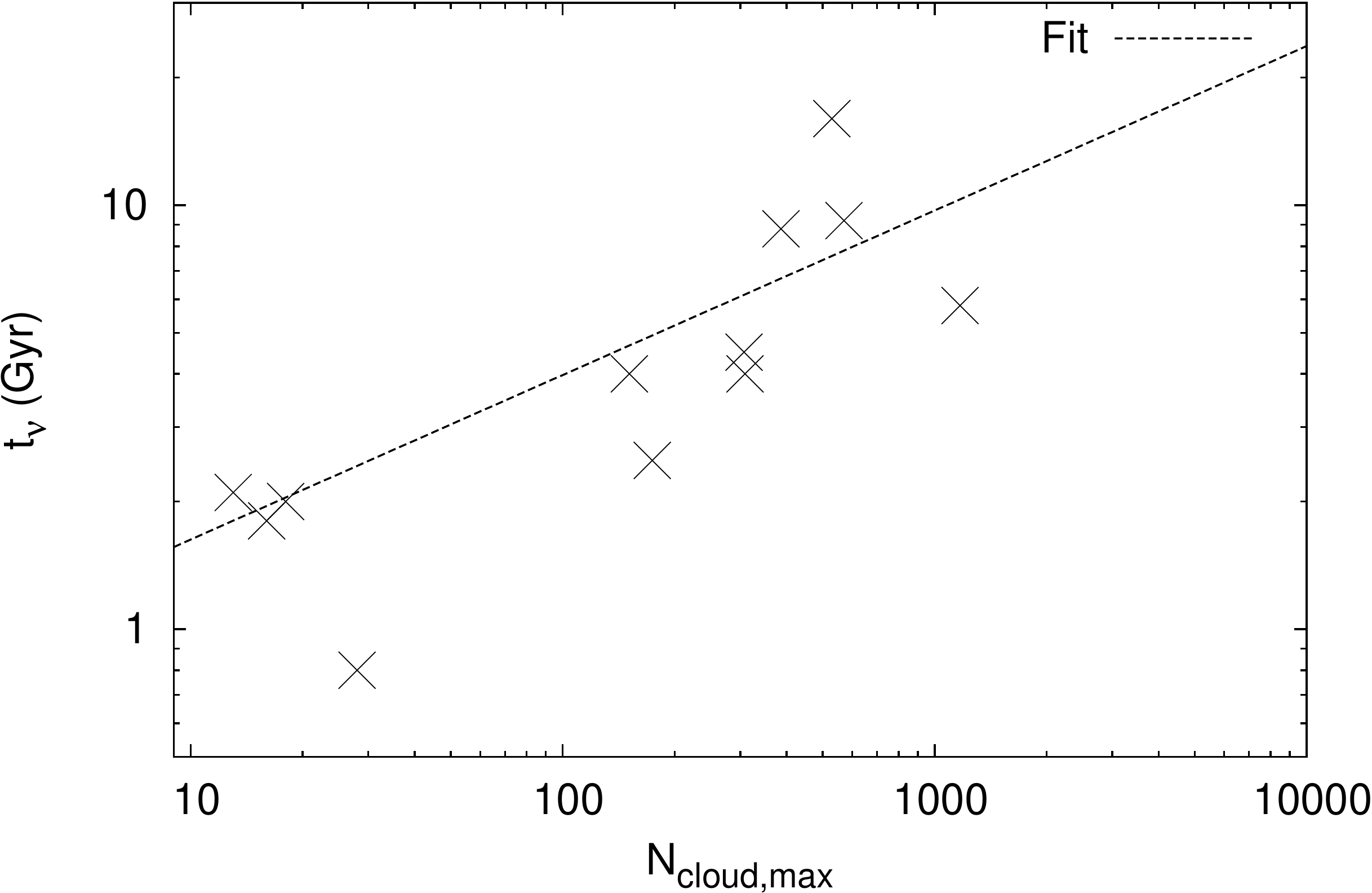}\\
\end{center}
\caption{\label{viscncloud} Correlation between peak number of clouds $(N_{\rm cloud,max})$ and viscous time-scales ($t_\nu$) for all models whose time-scale is given in Table~\ref{visctimes}. The plotted fit is $t_\nu = (0.67 {\rm~Gyr})(N_{\rm cloud,max})^{0.39}$.}
\end{figure}

Interestingly, despite the different disc properties, LowSoftC and LowSoftMW have similar viscous time-scales in both our numerical simulations and in this analysis. This is to be expected from equation \ref{correctvisceq}. We should expect the typical cloud mass to increase with the gas density and typical cloud radius, and so $M_{\rm cloud}/(\pi r_{\rm cloud}^2 \Sigma_{\rm g})$ should vary only weakly. Hence the viscous time-scale will primarily depend primarily on $h$ and $v_s$. This suggests that time-scales will not vary greatly for models beyond those simulated here - perhaps even of higher resolution. To quantify this, we note that there appears to be a correlation between the maximum number of clouds formed ($N_{\rm cloud,max}$) and the viscous time-scale (Fig.~\ref{viscncloud}). Performing a fit to a power-law $t_\nu \propto (N_{\rm cloud,max})^m$, we find a power index of $m=0.39\pm0.19$. This predicts a viscous time-scale of $t_\nu\sim23$ Gyr for $N_{\rm cloud,max}=10^4$, and $t_\nu\sim60$ Gyr for $N_{\rm cloud,max}=10^5$, although we caution that this is a purely empirical fitting and is not likely to be very accurate.

\section{Conclusions}

Previous estimates of the viscous time-scale suggest that the viscous time-scale for cloud-cloud collisions in a Milky-Way-like galaxy is large, with $t_\nu>1000$ Gyr. To test the hypothesis that the viscous time-scale is long, we performed simulations using the AP$^3$M SPH code HYDRA with cooling down to $10$ K and a dynamic temperature floor. The simulations fell into two sets of models: initially stable gaseous discs within a dark-matter haloes and stellar discs, and a gaseous spheres collapsing inside dark-matter haloes. These two models were chosen to bracket a wide range of stability. The viscous time-scale was measured by tracking clouds with a friends-of-friends algorithm, and determining the energy loss when clouds collided.

Although our cloud masses are larger than those found in other simulations, potentially due to insufficient resolution, a simple analysis suggests that we are resolving the wavelength of the most unstable mode. However, further instabilities (in particular, non-axisymmetric instabilities that we exclude from our analysis) may appear at higher resolutions, and while the inclusion of energy input from stellar feedback may not greatly alter the properties of clouds, it may contribute to cloud evaporation and affect their collisional behaviour by increasing their cross-section through heating.

Identifying clouds and interactions between clouds is still a difficult task, as clouds have complex structures and dynamics. The friends-of-friends algorithm often identifies clouds as merging and separating several times over a period that upon visual inspection appears to be a single interaction. Through a detailed examination of $10$ interaction events, we determined that each `real' interaction corresponds to $\sim4.5$ interactions found by our algorithms. This also complicated our estimates for $\eta=\Delta K_{\rm cloud}/K_{\rm cloud}$, the efficiency of energy loss per cloud interaction. We found that despite our low viscous time-scales, $\eta$ was not large, with $\eta\sim0.002$ per recorded interaction.

Most models from both sets of initial conditions collapsed into discs dominated by clumps of gas. The Milky Way models produced a more stable disc with a large number of small clouds, while the collapse models produced a highly unstable disc consisting of a small number of massive clumps. Despite this large disparity, the viscous time-scales were similar, with $t_\nu=4.5$ Gyr for LowSoftMW, and $t_\nu=0.8$ Gyr for LowSoftC. These values are much smaller than estimates using the formulation of \bellcitecom which overestimate the viscous time-scale by appearing to erroneously include inefficiency of cloud collisions twice. Removing this factor gives analytic estimates of $t_\nu=1.1$ Gyr for LowSoftMW and $t_\nu=17$ Gyr for LowSoftC. These values do not exactly coincide with our measured values as they are based on simple arguments that are particularly inaccurate for LowSoftC. However, they all agree with the general statement that viscosity due to cloud-cloud collisions is not negligible.

The scatter of $t_\nu$ across our models was quite small ($0.6$--$16.0$ Gyr), despite the range of cloud properties. Hence our viscous time-scales are applicable for a wider range of galaxies than those modelled here, although viscous time-scales will likely increase somewhat as resolution improves. For a simulation capable of resolving $10^5$ clouds, we predict a viscous time-scale of around $60$ Gyr, admittedly making the effect comparatively weak within a Hubble time, but nonetheless over an order of magnitude faster than previous estimates.

These results suggest that viscosity due to cloud-cloud collisions, while not dominant, does not have a completely negligible effect on the evolution of a galaxy. Although our models may underestimate the viscous time-scales due to resolution effects, it still appears that cloud-cloud viscosity is more significant than previously estimated. While numerical models of galaxies may be able to model this directly (as we do in this work), it may be necessary to include a cloud-cloud viscous term in analytical and semi-analytical models of disc evolution.

\section*{Acknowledgments}

We thank Larry Widrow for providing his NFW halo generator, Tobias Kauffman for providing his high entropy function 
definition, and our anonymous referee for suggestions that improved this paper. We also thank Brad Gibson for helpful discussions. The Friends-of-Friends code we used was downloaded from the University of Washington N-body shop {\tt 
http://www-hpcc.astro.washington.edu/}.  RJT is supported by a Discovery Grant from NSERC, the Canada Foundation for 
Innovation, the Nova Scotia Research and Innovation Trust and the Canada Research Chairs Program. Simulations were run on 
the CFI-NSRIT funded {\em St Mary's Computational 
Astrophysics Laboratory}.

\bibliographystyle{mnras}
\bibliography{willthackcloudcols}

\bsp

\label{lastpage}

\end{document}